\documentclass[10pt,a4paper,reqno]{amsart}
\usepackage{acronym}
\usepackage{amsfonts}
\usepackage{amsmath}
\usepackage{amssymb}
\usepackage{amsthm}
\usepackage{bm}
\usepackage{braket}
\usepackage{cite}
\usepackage{color}
\usepackage{geometry}
\usepackage{graphicx}
\usepackage{hyperref}
\usepackage{mathrsfs}
\usepackage{mathtools}
\usepackage{setspace}
\usepackage{stmaryrd}
\usepackage{subcaption}
\usepackage{tikz}

\usepackage{epsfig}
\usepackage{setspace}
\usepackage{booktabs}
\usepackage{threeparttable}
\usepackage{diagbox}
\usepackage{epstopdf}

\newgeometry{top=2cm,bottom=2cm,outer=1.5cm,inner=1.5cm}
\newcommand{\bse}{\begin{subequations}}
\newcommand{\ese}{\end{subequations}}

\numberwithin{equation}{section}

\title[Bound-state soliton and rogue wave solutions for the
sixth-order nonlinear Schr\"{o}dinger equation]{Bound-state soliton and rogue wave solutions for the sixth-order nonlinear Schr\"{o}dinger equation via inverse scattering transform method}
\author{Weiqi Peng}
\address[WP]{School of Mathematical Sciences, Shanghai Key Laboratory of Pure Mathematics and Mathematical Practice\\
East China Normal University \\ Shanghai 200241 \\ People's Republic of China}
\author{Yong Chen$^*$}
\address[YC]{School of Mathematical Sciences, Shanghai Key Laboratory of Pure Mathematics and Mathematical Practice\\
East China Normal University \\ Shanghai 200241 \\ People's Republic of China}
\address[YC]{College of Mathematics and Systems Science \\ Shandong University of Science and Technology \\ Qingdao 266590 \\ People's Republic of China}
\email{ychen@sei.ecnu.edu.cn($^*$Corresponding author).}

\begin{document}

\begin{abstract}
In this work,  inverse scattering
transform for the
sixth-order nonlinear Schr\"{o}dinger equation with both
zero and nonzero boundary conditions at infinity is given, respectively. For the case of zero nonzero boundary conditions, in terms of the Laurent's series and generalization of the residue theorem, the  bound-state soliton is derived. For nonzero boundary conditions, using the robust inverse scattering transform, we present a matrix Riemann-Hilbert problem of the sixth-order nonlinear Schr\"{o}dinger equation. Then based on the obtained Riemann-Hilbert problem, the rogue wave and breather wave solutions are derived through a modified Darboux transformation. Besides, according to some appropriate parameters choices, several graphical analyses are provided to discuss the dynamical behaviors of the bound-state soliton and rogue wave solutions, and analyse how the higher-order terms  affect the
bound-state soliton and rogue wave.
\\
\\
\emph{Key words:} Inverse scattering transform; Sixth-order nonlinear Schr\"{o}dinger equation;  Bound-state soliton; Rogue wave.\\
\end{abstract}

\maketitle

\section{Introduction}
The solutions of nonlinear evolution equations with zero boundary conditions (ZBCs) and nonzero boundary conditions (NZBCs) play a vital role in describing some relevant phenomena, and more
and more researchers are interested in studying this topic. In the development of the past few decades, many techniques have been
provided to find these solutions \cite{Wang1,Wang2,Wang3,Wang4}.
Of which, the inverse scattering transform (IST)
method is the most powerful tool for analyzing  initial value problems of integrable systems in
soliton theory. The method was presented for the first time by Gardner et al. in 1967 for the KdV equation\cite{Wang14}.
In a general way, the classical IST method was based on the Gel'fand-Levitan-Marchenko(GLM) integral
equations. After that, Zakharov et al. properly simplified the IST method through developing a Riemann-Hilbert formulation\cite{Wang15}. Undergoes decades of development, the researches of Riemann-Hilbert formulation have made many successful  progresses in the area of integrable systems, and it is
still a hot topic today\cite{Wang16,Wang18,Gengxiang1,Gengxiang2,Gengxiang3,Gengxiang4,Wang19,Jenkins,Deift,Biondini1,Wang21}. Especially, in recent years, using the Riemann-Hilbert method to research the bound-state soliton and rogue wave solutions  has attracted many attentions.

The bound-state solitons, called multiple-poles soliton solutions, are generated when two or more fundamental solitons coexist with the same velocity and the same position. The multiple-poles soliton solutions for focusing nonlinear Schr\"{o}dinger (NLS) equation was first derived by Zakharov and Shabat\cite{ZhangXiaoEn15}. Since then, the multiple-poles solitons for various nonlinear integrable equations have be obtained, such as the sine-Gordon equation \cite{ZhangXiaoEn16,ZhangXiaoEn17}, the modified
KdV equation\cite{ZhangXiaoEn18}, the complex
modified KdV equation\cite{bs1}, Wadati-Konno-Ichikawa equation \cite{bs2},  Sasa-Satsuma
equation\cite{lin1,Yang1}. As well as,  asymptotic of multiple-pole solitons was analysed in \cite{JNS1,ZhangXiaoEn}. Generally,  the IST method with GLM equation results in a complicated calculation
for finding the bound-state solitons\cite{ZhangXiaoEn17, ZhangXiaoEn18}, since some complicate limits need to be settled. However, through using the Laurent's series and generalization of the residue theorem\cite{bs1,bs2,bs3}, the Riemann-Hilbert problem(RHP) with multiple-poles can be solved directly, and the bound-state solitons can be expressed compactly.

Rogue waves have been constantly observed in various fields, which appear suddenly and disappear without trace, mainly possessing a high peak. In the study of the NLS equation, Peregrine first derived a kind of rational solution to describe rogue wave phenomena\cite{Peng12}. After that, the rogue waves have attracted the attention of more and more researchers. During this period, a lot of technologies have  been developed to study the rogue waves, containing the Wronskian technique, the Darboux transformation (DT) method,  the KP reduction method, etc. \cite{Peng21,Peng24,Peng26,Peng27,Peng32,Chen6}. However, the standard IST is difficult to generate the rogue wave solution of the nonlinear integrable equations because of special regularity of the singular point.  Until a wonderful and powerful approach, named robust inverse scattering transform, was Pioneered by Bilman and Miller to research higher-order rogue wave solutions for the focusing NLS equation\cite{Cheng10}. Suddenly,
the robust inverse scattering transform has been employed to study the rogue waves for some nonlinear integrable model, including the Hirota equation\cite{Liu35}, fifth-order nonlinear Schr\"{o}dinger equation\cite{Liu36}, generalized nonlinear
Schr\"{o}dinger equation\cite{Liu37}, quartic nonlinear Schr\"{o}dinger equation\cite{Liu38}.

Nowadays, the investigation of high order NLS equations has attracted more and more attention, since the NLS equation with high-order dispersion, self-steepening, high-order nonlinearity and self-frequency shift can be better used to describe the propagation of ultrashort pulses in optical fibers\cite{He-PRE}. For this reason,  in this paper, we mainly focus on the sixth-order nonlinear Schr\"{o}dinger(SONLS) equation\cite{Akhmediev}
\begin{align}\label{1}
iq_{t}+\frac{1}{2}q_{xx}+q|q|^{2}+\delta K(q)=0,
\end{align}
where
\begin{align}\label{2}
K(q)=&q_{xxxxxx}+q^{2}\left[60|q_{x}|^{2}q^{\ast}+50q_{xx}(q^{\ast})^{2}+2q_{xxxx}^{\ast}\right]\notag\\
&+q\left[12q^{\ast}q_{xxxx}+18q_{x}^{\ast}q_{xxx}+8q_{x}q_{xxx}^{\ast}+70\left(q^{\ast}\right)^{2}q_{x}^{2}+22|q_{xx}|^{2}\right]\notag\\
&+10q_{x}\left[3q^{\ast}q_{xxx}+5q_{x}^{\ast}q_{xx}+2q_{x}q_{xx}^{\ast}\right]
+10q^{3}\left[2q^{\ast}q_{xx}^{\ast}+\left(q_{x}^{\ast}\right)^{2}\right]\notag\\
&+20q^{\ast}q_{xx}^{2}+20q|q|^{6},
\end{align}
of which $\delta$ is a arbitrary real constant which represents the
strength of higher-order nonlinear effects. $q(x, t)$ is a complex valued function with the real variables $x$ and $t$.
Eq.\eqref{1} can be applied to depict the propagation of subpicosecond and femtosecond pulses in the fibers. In terms of the Hirota method, the soliton solutions of Eq.\eqref{1} were given in \cite{Yue48,Yue50}. The DT method is used to study the breather-to-soliton transition for the SONLS Eq.\eqref{1} in \cite{Yue49}. Recently, through the generalized DT method, the Modulation instability and rogue wave solutions for Eq.\eqref{1} were studied in \cite{Yue}.  Eq.\eqref{1} subjects to the following Lax pair\cite{Akhmediev}
\begin{align}\label{3.1}
\phi_{x}=U\phi=(ik\sigma_{3}+Q)\phi,\qquad \phi_{t}=V\phi=\sum_{m=0}^{6}ik^{m}V_{m}\phi,
\end{align}
where
\begin{align}\label{4}
Q=\left(\begin{array}{cc}
    0  &  iq^{\ast}\\
    iq &  0\\
\end{array}\right),\qquad \sigma_{3}=\left(\begin{array}{cc}
    1  &  0\\
    0 &  -1\\
\end{array}\right), \qquad V_{m}=\left(\begin{array}{cc}
    A_{m}  &  B_{m}^{\ast}\\
   B_{m} &  -A_{m}\\
\end{array}\right),
\end{align}
with
\begin{align}\label{5}
A_{0}=&-\frac{1}{2}|q|^{2}-10\delta|q|^{6}-5\delta\left[q^{2}(q_{x}^{\ast})^{2}+(q^{\ast})^{2}q_{x}^{2}\right]
-10\delta|q|^{2}(qq_{xx}^{\ast}+q^{\ast}q_{xx})\notag\\
&-\delta|q_{xx}|^{2}+\delta(q_{x}q_{xxx}^{\ast}
+q_{x}^{\ast}q_{xxx}-q^{\ast}q_{xxxx}-qq_{xxxx}^{\ast}),\notag\\
A_{1}=&12i\delta|q|^{2}(q_{x}q^{\ast}-q_{x}^{\ast}q)+2i\delta(q_{x}q_{xx}^{\ast}-q_{x}^{\ast}q_{xx}+q^{\ast}q_{xxx}-q_{xxx}^{\ast}q),\notag\\
A_{2}=&1+12\delta|q|^{4}-4\delta|q_{x}|^{2}+4\delta(q_{xx}^{\ast}q+q_{xx}q^{\ast}),\ A_{3}=8i\delta(qq_{x}^{\ast}-q^{\ast}q_{x}),\notag\\
A_{4}=&-16\delta|q|^{2}, \ A_{5}=0,\ A_{6}=32\delta, \ B_{2}=-24i\delta|q|^{2}q_{x}-4i\delta q_{xxx},\ B_{4}=16i\delta q_{x},\notag\\
B_{0}=&\frac{i}{2}q_{x}+i\delta q_{xxxxx}+10i\delta(qq_{x}^{\ast}q_{xx}+qq_{xx}^{\ast}q_{x}+|q|^{2}q_{xxx}+3|q|^{4}q_{x}+q_{x}|q_{x}|^{2}+2q^{\ast}q_{x}q_{xx}),
\notag\\
B_{1}=&q+12\delta q^{\ast}q_{x}^{2}+16\delta|q|^{2}q_{xx}+4\delta q^{2}q_{xx}^{\ast}+2\delta q_{xxxx}+12\delta|q|^{4}q+8\delta q|q_{x}|^{2},\notag\\
B_{3}=&-16\delta|q|^{2}q-8\delta q_{xx},\ B_{5}=32\delta q,\ B_{6}=0,
\end{align}
where $k$ is the spectrum parameter, the superscript $\ast$ means the complex conjugate,
and the function $\phi$ is a $2\times 2$  eigenfunction matrix.
The primary purpose of the present paper is to discuss the bound-state soliton and rogue wave solutions for the SONLS equation through inverse scattering transform method, and analyse how the higher-order terms  affect the
bound-state soliton and rogue wave by  adjusting the value of $\delta$.

The outline of this paper is organized as follows: In section 2, we construct the reflection-less RHP  with one higher-order poles, and obtain the multiple bound-state soliton solutions for the SONLS equation with ZBCs. Besides, the compression effects on bound-state soliton for different $\delta$ are given. In section 3, we establish the RHP for the SONLS equation with NZBCs using the robust inverse scattering transform. Then the RHP can be solved by a modified Darboux transformation, and the exact breather wave and rogue wave solutions are further presented for the SONLS equation. Similarly, we also compare the effect of the higher order terms $\delta$ on the rogue wave. Finally, some conclusions are given in the last section.

\section{The IST with ZBCs and bound-state soliton}
In this section, we will seek the bound-state soliton $q(x, t)$ for the SONLS Eq.\eqref{1} with  ZBCs at infinity given by
\begin{align}\label{5a}
\lim_{x\rightarrow \pm \infty}q(x, t)=0.
\end{align}
In what follows, we will present the IST and bound-state soliton for Eq.\eqref{1} with ZBCs by solving a matrix RHP.

\subsection{The construction of the RHP with ZBCs}
Let $x \rightarrow \pm \infty$,  the Lax pair \eqref{3.1} under the boundary \eqref{5a} can be changed into
\begin{align}\label{6a}
\phi_{x}=U_{0}\phi=ik\sigma_{3}, \qquad \phi_{t}=V_{0}\phi=(k+32\delta k^{5})U_{0}\phi,
\end{align}
which admits the fundamental matrix solution $\phi_{bg}(x, t; k)$, given by
\begin{align}\label{7a}
\phi^{bg}(x, t; k)=e^{i\Theta(x, t; k)\sigma_{3}},\qquad \Theta(x, t; k)=k[x+(k+32\delta k^{5})t].
\end{align}
Then, we can find the following Jost solutions $\phi_{\pm}(x, t,; k)$
\begin{align}\label{8a}
\phi_{\pm}(x, t; k)\rightarrow e^{i\Theta(x, t; k)\sigma_{3}}, \quad \mbox{as} \quad x\rightarrow \pm \infty.
\end{align}
Further, the modified Jost solutions $\mu_{\pm}(x, t; k)$ are taken as
\begin{align}\label{9a}
\mu_{\pm}(x, t; k)=\phi_{\pm}(x, t; k)e^{-i\Theta(x, t; k)\sigma_{3}},
\end{align}
which results in $\mu_{\pm}(x, t; k)\rightarrow \mathbb{I}$ as $x\rightarrow \pm\infty$. Then,  the following Volterra integral equations are satisfied
\begin{align}\label{10a}
\begin{cases}
\mu_{-}(x, t;k)=\mathbb{I}+\int_{-\infty}^{x}\exp\left[ik\sigma_{3}\left(x-y\right)\right]Q(y,t)\mu_{-}(y,t;k)
\exp\left[ik\sigma_{3}\left(y-x\right)\right]\mathrm{d}y, \\
\mu_{+}(x, t;k)=\mathbb{I}-\int_{x}^{+\infty}\exp\left[ik\sigma_{3}\left(x-y\right)\right]Q(y,t)\mu_{+}(y,t;k)
\exp\left[ik\sigma_{3}\left(y-x\right)\right]\mathrm{d}y.
\end{cases}
\end{align}
Let $\mathbb{C}_{\pm}= \left\{k\in \mathbb{C} | \mbox{Im}k\gtrless 0\right\}$ (see Fig. 1). It is not hard to find that the columns $\mu_{+,1}$ and $\mu_{-,2}$ are analytic in $\mathbb{C}_{+}$, and continuously extended to $\mathbb{C}_{+}\cup \mathbb{R}$. $\mu_{-,1}$ and $\mu_{+,2}$ are analytic in $\mathbb{C}_{-}$, and continuously extended to $\mathbb{C}_{-}\cup \mathbb{R}$.

\centerline{\begin{tikzpicture}[scale=1.5]
\path [fill=gray] (-2.5,0) -- (-0.5,0) to
(-0.5,2) -- (-2.5,2);
\path [fill=gray] (-4.5,0) -- (-2.5,0) to
(-2.5,2) -- (-4.5,2);
\draw[-][thick](-4.5,0)--(-2.5,0);
\draw[fill] (-2.5,0) circle [radius=0.03];
\draw[->][thick](-2.5,0)--(-0.5,0)node[above]{$\mbox{Re}k$};
\draw[->][thick](-2.5,1)--(-2.5,2)node[right]{$\mbox{Im}k$};
\draw[-][thick](-2.5,1)--(-2.5,0);
\draw[-][thick](-2.5,0)--(-2.5,-1);
\draw[-][thick](-2.5,-1)--(-2.5,-2);
\draw[fill] (-2.5,-0.3) node[right]{$0$};
\draw[fill] (-1.7,0.8) circle [radius=0.03] node[right]{$k_{n}$};
\draw[fill] (-1.7,-0.8) circle [radius=0.03] node[right]{$k^{*}_{n}$};
\end{tikzpicture}}
\noindent { \small \textbf{Figure 1.} (Color online) Distribution of the discrete spectrum and the contours for the RHP on complex $k$-plane, Region $\mathbb{C}_{+}$ (gray region), region $\mathbb{C}_{-}$ (white region).}\\

In fact, the Jost solutions
$\phi_{\pm}(x, t; k)$ are the simultaneous solutions for the Lax pair \eqref{3.1}. Therefore, $\phi_{\pm}(x, t; k)$ have following linear relation by  the constant scattering matrix $S(k)=(s_{i j} (k))_{2\times 2}$
\begin{align}\label{11a}
\phi_{+}(x, t; k)=\phi_{-}(x, t; k)S(k), \quad k\in \mathbb{R},
\end{align}
or
\begin{align}\label{11b}
\mu_{+}(x, t; k)=\mu_{-}(x, t; k)e^{i\Theta\sigma_{3}}S(k)e^{-i\Theta\sigma_{3}}, \quad k\in \mathbb{R},
\end{align}
where $S(k)=\sigma_{2}S^{\ast}(k^{\ast})\sigma_{2}, s_{11}(k)=s_{22}^{\ast}(k^{\ast}), s_{12}(k)=-s_{21}^{\ast}(k^{\ast})$, and $\sigma_{2}=\left(\begin{array}{cc}
    0  &  -i\\
  i &  0\\
\end{array}\right)$. Due to the scattering coefficients can be expressed as the following Wronskians determinant form
\begin{align}\label{12a}
s_{11}(k)=Wr(\phi_{+,1},\phi_{-,2}), \quad s_{12}(k)=Wr(\phi_{+,2},\phi_{-,2}),\notag\\
s_{21}(k)=Wr(\phi_{-,1},\phi_{+,1}), \quad s_{22}(k)=Wr(\phi_{-,1},\phi_{+,2}),
\end{align}
we find that $s_{11}$ can be  analytic continuation to
$\mathbb{C}_{+}$, and analogously $s_{22}$ is analytic in $\mathbb{C}_{-}$. In addition, $s_{11}$ , $s_{22}\rightarrow 1$  as $k\rightarrow\infty$ in $\mathbb{C}_{+}$ and $\mathbb{C}_{-}$, respectively.

In order to construct a RHP for the inverse spectral problem, we consider the following sectionally meromorphic matrices
\begin{align}\label{25a}
M_{+}(x, t, k)=(\frac{\mu_{+,1}}{s_{11}},\mu_{-,2}),\qquad M_{-}(x, t, k)=(\mu_{-,1},\frac{\mu_{+,2}}{s_{22}}),
\end{align}
where superscripts $\pm$ denote analyticity in $\mathbb{C}_{+}$ and $\mathbb{C}_{-}$, respectively. Naturally, a matrix Riemann-Hilbert problem is presented:

\noindent \textbf{Riemann-Hilbert Problem 1}  \emph{
$M(k; x, t)$ solve the following RHP:
\begin{align}\label{26a}
\left\{
\begin{array}{lr}
M(k; x, t)\ \mbox{is analytic in} \ \mathbb{C}\setminus \mathbb{R},\\
M_{+}(k; x, t)=M_{-}(x, t, k)G(x, t, k), \qquad k\in \mathbb{R},\\
M(k; x, t)\rightarrow \mathbb{I},\qquad k\rightarrow \infty,
  \end{array}
\right.
\end{align}
of which the jump matrix $G(x, t, k)$ is
\begin{align}\label{27a}
G(x, t, k)=\left(\begin{array}{cc}
    1+|r(k)|^{2}  &  e^{2i\Theta(x, t, k)}r^{\ast}(k)\\
  e^{-2i\Theta(x, t, k)}r(k) &  1\\
\end{array}\right),
\end{align}
where $r(k)=\frac{s_{21}(k)}{s_{11}(k)}$.}

Recalling the symmetry properties of the Jost eigenfunctions and scattering coefficients, we have
$M_{+}(k)=\sigma_{2}M_{-}^{\ast}(k^{\ast})\sigma_{2}$. Taking
\begin{align}\label{36a}
M(x, t; k)=\mathbb{I}+\frac{1}{k}M^{(1)}(x, t; k)+O(\frac{1}{k^{2}}),\qquad k\rightarrow \infty,
\end{align}
then the potential $q(x, t)$ of the SONLS equation with ZBCs is given by
\begin{align}\label{38a}
q(x, t)=2M_{21}^{(1)}(x, t, k)=\lim_{k\rightarrow\infty}2kM_{21}(x, t, k).
\end{align}

\subsection{Bound-state soliton  with one higher-order
pole}
In general, there are exactly discrete spectral points $k$ in $\mathbb{C}_{+}$ which make $s_{11}(k)=0$ and those discrete  spectral points  in $\mathbb{C}_{-}$ lead to $s_{22}(k)=0$. Without considering simple poles, we here
assume that $s_{11}(k)$ has $N$ higher-order poles $k_{n}$, $n=1, 2,\cdots, N,$ in $\mathbb{C}_{+}$, that means
\begin{align}\label{17b}
s_{11}(k)=(k-k_{1})^{n_{1}}(k-k_{2})^{n_{2}}\times\cdots\times(k-k_{N})^{n_{N}}s^{(0)}_{11}(k),\notag\\
s_{22}(k)=(k-k_{1}^{\ast})^{n_{1}}(k-k_{2}^{\ast})^{n_{2}}\times\cdots\times(k-k_{N}^{\ast})^{n_{N}}s^{(0)}_{22}(k),
\end{align}
where $s^{(0)}_{11}(k)=s^{(0)\ast}_{22}(k^{\ast})\neq 0$ for all $k\in\mathbb{C}_{+}$.
The  symmetric relation  of the scattering matrix yields
\begin{align}\label{18a}
s_{11}(k_{n})=s_{22}^{\ast}(k_{n}^{\ast})=0.
\end{align}
Thus, the corresponding discrete spectral point set is
\begin{align}\label{19a}
\Gamma=\left\{k_{n}, k_{n}^{\ast}\right\}_{n=1}^{N},
\end{align}
whose distributions are shown in Fig. 1.

To obtain the explicit soliton solutions, we consider the reflectionless potential of the SONLS  equation with ZBCs i.e., $r(k)=0$. As a matter of convenience, we will show the case of one higher-order pole. This case implies  $s_{11}(k)$ has one $N$th order zero point on the upper half plane, i.e.,  $s_{11}(k)=(k-k_{0})^{N}s^{(0)}_{11}(k)(\mbox{Im} k>0, N>1, s^{(0)}_{11}(k_{0})\neq 0)$.
As well, we know that
$M_{11}(x, t, k)$ has one $N$th order pole at $k=k_{0}$, and $M_{12}(x, t, k)$
has one $N$th order pole at $k=k_{0}^{\ast}$. According to normalization condition of matrix $M(x, t, k)$, we can write the RHP in the following form
\begin{align}\label{39a}
M_{11}(x, t, k)=1+\sum_{n=1}^{N}\frac{F_{n}(x, t)}{(k-k_{0})^{n}},\qquad M_{12}(x, t, k)=\sum_{n=1}^{N}\frac{G_{n}(x, t)}{(k-k_{0}^{\ast})^{n}}.
\end{align}
Simultaneously, defining
\begin{align}\label{40a}
&e^{-2i\Theta}=\sum_{s=0}^{+\infty}f_{s}(x,t)(k-k_{0})^{s}, \ M_{12}(x,t,k)=\sum_{s=0}^{+\infty}g_{s}(x,t)(k-k_{0})^{s}, \notag\\
&e^{2i\Theta}=\sum_{s=0}^{+\infty}f^{\ast}_{s}(x,t)(k-k^{\ast}_{0})^{s}, \ M_{11}(x,t,k)=\sum_{s=0}^{+\infty}\varrho_{s}(x,t)(k-k_{0}^{\ast})^{s},\notag\\
&r(k)=r_{0}(k)+\sum_{m=1}^{N}\frac{r_{m}}{(k-k_{0})^{m}},\ r^{\ast}(k^{\ast})=r^{\ast}_{0}(k^{\ast})+\sum_{m=1}^{N}\frac{r^{\ast}_{m}}{(k-k_{0}^{\ast})^{m}},
\end{align}
where
\begin{align}\label{41a}
&f_{s}(x,t)=\lim_{k\rightarrow k_{0}}\frac{1}{s!}\frac{\partial^{s}}{\partial k^{s}}e^{-2i\Theta}, \ g_{s}(x, t)=\lim_{k\rightarrow k_{0}}\frac{1}{s!}\frac{\partial^{s}}{\partial k^{s}}M_{12}(x,t,k),\notag\\
&\varrho_{s}(x, t)=\lim_{k\rightarrow k_{0}^{\ast}}\frac{1}{s!}\frac{\partial^{s}}{\partial k^{s}}M_{11}(x,t,k), \
r_{m}=\lim_{k\rightarrow k_{0}}\frac{1}{(N-m)!}\frac{\partial^{N-m}}{\partial k^{N-m}}[(k-k_{0})^{N}r(k)], \notag\\
&s=0,1,2,3,\cdots,  m=1,2,3,\cdots, N,
\end{align}
and $r_{0}(k)$ is analytic on the upper half plane.

According to the RHP \textbf{1}, \eqref{39a}, and \eqref{40a}, we can collect the
associated coefficients of $(k-k_{0})^{-n}$ and $(k-k_{0}^{\ast})^{-n}$, which leads to
\begin{align}\label{41.1a}
&F_{n}(x, t)=\sum_{j=n}^{N}\sum_{s=0}^{j-n}r_{j}f_{j-n-s}(x, t)g_{s}(x, t),\notag\\
&G_{n}(x, t)=-\sum_{j=n}^{N}\sum_{s=0}^{j-n}r^{\ast}_{j}f^{\ast}_{j-n-s}(x, t)\varrho_{s}(x, t),\ n=1, 2, \ldots, N.
\end{align}
Furthermore, putting \eqref{39a} into $g_{s}, \varrho_{s}$ given in \eqref{41a}, we obtain following results
\begin{align}\label{41.2a}
&g_{s}(x, t)=\sum_{n=1}^{N}\left(\begin{array}{c}
    n+s-1\\
    s\\
\end{array}\right)\frac{(-1)^{s}G_{n}}{(k_{0}-k_{0}^{\ast})^{n+s}},\ s=0, 1, 2,\ldots \notag\\
&\varrho_{s}(x, t)=\left\{
\begin{array}{lr}
1+\sum_{n=1}^{N}\frac{F_{n}}{(k_{0}^{\ast}-k_{0})^{n}},\ s=0,\\
\sum_{n=1}^{N}\left(\begin{array}{c}
    n+s-1\\
    s\\
\end{array}\right)\frac{(-1)^{s}F_{n}}{(k_{0}^{\ast}-k_{0})^{n+s}},\ s=1, 2, 3,\ldots
  \end{array}
\right.
\end{align}
The substitution of \eqref{41.2a} into \eqref{41.1a} results in
\begin{align}\label{41.3a}
&F_{n}(x, t)=\sum_{j=n}^{N}\sum_{s=0}^{j-n}\sum_{p=1}^{N}\left(\begin{array}{c}
    p+s-1\\
    s\\
\end{array}\right)\frac{(-1)^{s}r_{j}f_{j-n-s}(x, t)G_{p}}{(k_{0}-k_{0}^{\ast})^{p+s}},\notag\\
&G_{n}(x, t)=-\sum_{j=n}^{N}r^{\ast}_{j}f^{\ast}_{j-n}(x, t)\notag\\
&-\sum_{j=n}^{N}\sum_{s=0}^{j-n}\sum_{p=1}^{N}\left(\begin{array}{c}
    p+s-1\\
    s\\
\end{array}\right)\frac{(-1)^{s}r^{\ast}_{j}f^{\ast}_{j-n-s}(x, t)F_{p}}{(k_{0}^{\ast}-k_{0})^{p+s}},\ n=1, 2, \ldots, N.
\end{align}
Ultimately, the following theorem can be summarized:

\noindent \textbf{Theorem 1}  \emph{
With the ZBCs at infinity given by \eqref{5a}, the $N$th order bound-state
soliton of the  SONLS equation is
\begin{align}\label{42a}
q=2\left[1-\frac{\det(\mathbb{I}+\Omega\Omega^{\ast}+|\chi^{\ast}\rangle\langle P_{0}|)}{\det (\mathbb{I}+\Omega\Omega^{\ast})}\right],
\end{align}
where $\langle P_{0}|=[1, 0, 0, \cdots, 0]_{1\times N},$ and
\begin{align}\label{43a}
&\Omega=[\Omega_{np}]_{N\times N}=\left[\sum_{j=n}^{N}\sum_{s=0}^{j-n}\left(\begin{array}{c}
    p+s-1\\
  s\\
\end{array}\right)\frac{(-1)^{s}r_{j}f_{j-n-s}}{(k_{0}-k_{0}^{\ast})^{p+s}}\right], \notag\\
& |\chi\rangle=[\chi_{1}, \chi_{2}, \chi_{3}, \cdots, \chi_{N}]^{T},\
\chi_{n}(x,t)=-\sum_{j=n}^{N}r_{j}^{\ast}f_{j-n}^{\ast},\qquad n, p=1, 2, 3,\cdots, N.
\end{align}
}
\noindent \textbf{Proof}  \emph{
Defining $| F\rangle=[F_{1}, F_{2},\ldots, F_{N}]^{T}$, $| G\rangle=[G_{1}, G_{2},\ldots, G_{N}]^{T}$, we can rewrite the \eqref{41.3a} as
\begin{align}\label{43.1a}
| F\rangle=\Omega| G\rangle, \ | G\rangle=| \chi\rangle-\Omega^{\ast}| F\rangle.
\end{align}
Then, we have
\begin{align}\label{43.2a}
| F\rangle=\Omega(\mathbb{I}+\Omega^{\ast}\Omega)^{-1}| \chi\rangle, \ | G\rangle=(\mathbb{I}+\Omega^{\ast}\Omega)^{-1}| \chi\rangle.
\end{align}
Substituting \eqref{43.2a} into \eqref{39a}, we obtain
\begin{align}\label{43.3a}
&M_{11}=\frac{\det (\mathbb{I}+\Omega^{\ast}\Omega+| \chi\rangle\langle P(k)|\Omega)}{\det(\mathbb{I}+\Omega^{\ast}\Omega)},\notag\\
&M_{12}=\frac{\det (\mathbb{I}+\Omega^{\ast}\Omega+| \chi\rangle\langle P^{\ast}(k^{\ast})|)}{\det(\mathbb{I}+\Omega^{\ast}\Omega)}-1,
\end{align}
where $\langle P(k)|=[\frac{1}{k-k_{0}}, \frac{1}{(k-k_{0})^{2}}, \cdots, \frac{1}{(k-k_{0})^{N}}]$. In terms of the symmetry relation for matrix $M$ and \eqref{38a}, the Theorem 1 can eventually be proved.
}

Taking $N=2$, then $k=k_{0}$ is a second-order zero point of $s_{11}$, $r(k)$ can be rewritten as
\begin{align}\label{46a}
r(k)=r_{0}(k)+\frac{r_{1}}{k-k_{0}}+\frac{r_{2}}{(k-k_{0})^{2}},
\end{align}
and elements of matrix $\Omega$ are
\begin{align}\label{47a}
&\Omega_{11}=\frac{r_{1}f_{0}}{k_{0}-k_{0}^{\ast}}+\frac{r_{2}f_{1}}{k_{0}-k_{0}^{\ast}}-\frac{r_{2}f_{0}}{(k_{0}-k_{0}^{\ast})^{2}},\notag\\
&\Omega_{12}=\frac{r_{1}f_{0}}{(k_{0}-k_{0}^{\ast})^{2}}+\frac{r_{2}f_{1}}{(k_{0}-k_{0}^{\ast})^{2}}-\frac{2r_{2}f_{0}}{(k_{0}-k_{0}^{\ast})^{3}},\notag\\
&\Omega_{21}=\frac{r_{2}f_{0}}{k_{0}-k_{0}^{\ast}}, \
\Omega_{22}=\frac{r_{2}f_{0}}{(k_{0}-k_{0}^{\ast})^{2}},
\end{align}
$|\chi\rangle$ is a column vector defined by
\begin{align}\label{48a}
\chi_{1}=-r_{1}^{\ast}f_{0}^{\ast}-r_{2}^{\ast}f_{1}^{\ast},\qquad \chi_{2}=-r_{2}^{\ast}f_{0}^{\ast},
\end{align}
and $\langle P_{0}|=[1, 0]$. For convenience, we have set $r_{1}=r_{2}=1$ and $k_{0}=a+bi$, in terms of Theorem \textbf{1}, the second order bound-state soliton solution of the SONLS
equation is
\begin{align}\label{49a}
q(x,t)=\frac{32b^{3}\Delta_{1}e^{-2i\nu_{1}}+32b^{3}\Delta_{2}e^{-2i\nu_{2}}}{(\Delta_{3}+256b^{8}e^{4b\nu_{3}})e^{4b\nu_{3}}},
\end{align}
where
\begin{align}\label{50a}
&\nu_{1}=(a+bi)((32b^{5}\delta i +160ab^{4}\delta -320a^{2}b^{3}\delta i-320a^{3}b^{2}\delta \notag\\
&+160a^{4}b \delta i+32a^{5}\delta +bi+a)t+x),\notag\\
&\nu_{2}=(-480a^{4}b^{2}\delta +480a^{2}b^{4}\delta -32b^{6}\delta +576a^{5}b\delta i-1920a^{3}b^{3}\delta i+576ab^{5}\delta i \notag\\
&+32a^{6}\delta -b^{2} +6abi+a^{2})t+(a+3bi)x,\notag\\
&\nu_{3}=(192a^{5}\delta -640a^{3}b^{2}\delta +192ab^{4}\delta +2a)t+x,\notag\\
&\Delta_{1}=(-6144a^{5}b^{5}\delta i+61440a^{3}b^{7}\delta i-30720ab^{9}\delta i+30720a^{4}b^{6}\delta -61440a^{2}b^{8}\delta \notag\\
&+6144b^{10}\delta -64ab^{5} i+64b^{6})t-32b^{5}ix+16b^{5},\notag\\
&\Delta_{2}=(384a^{5}b\delta i-3840a^{3}b^{3}\delta i+1920ab^{5}\delta i+1920a^{4}b^{2}\delta -3840a^{2}b^{4}\delta \notag\\
&+384b^{6}\delta +4ab i+4b^{2})t+2bix-2i+b,\notag\\
&\Delta_{3}=(9437184a^{10}b^{6}\delta^{2}+47185920a^{8}b^{8}\delta^{2}+94371840a^{6}b^{10}\delta^{2}+94371840a^{4}b^{12}
\delta^{2}\notag\\
&+47185920a^{2}b^{14}\delta^{2}+9437184b^{16}\delta^{2}+196608a^{6}b^{6}\delta-983040a^{4}b^{8}
\delta-983040a^{2}b^{10}\delta\notag\\
&+196608b^{12}\delta+1024a^{2}b^{6}+1024b^{8})t^{2}+(98304a^{5}b^{6}\delta-983040a^{3}b^{8}\delta+491520ab^{10}\delta\notag\\
&+1024ab^{6})xt+(245760a^{4}b^{7}\delta
-491520a^{2}b^{9}\delta+49152b^{11}\delta-49152a^{5}b^{5}\delta+491520a^{3}b^{7}\delta\notag\\
&-245760ab^{9}\delta+512b^{7}-512ab^{5})t+256b^{6}x^{2}+64b^{6}-256b^{5}x+96b^{4}.
\end{align}

\qquad
{\rotatebox{0}{\includegraphics[width=3.1cm,height=3.1cm,angle=0]{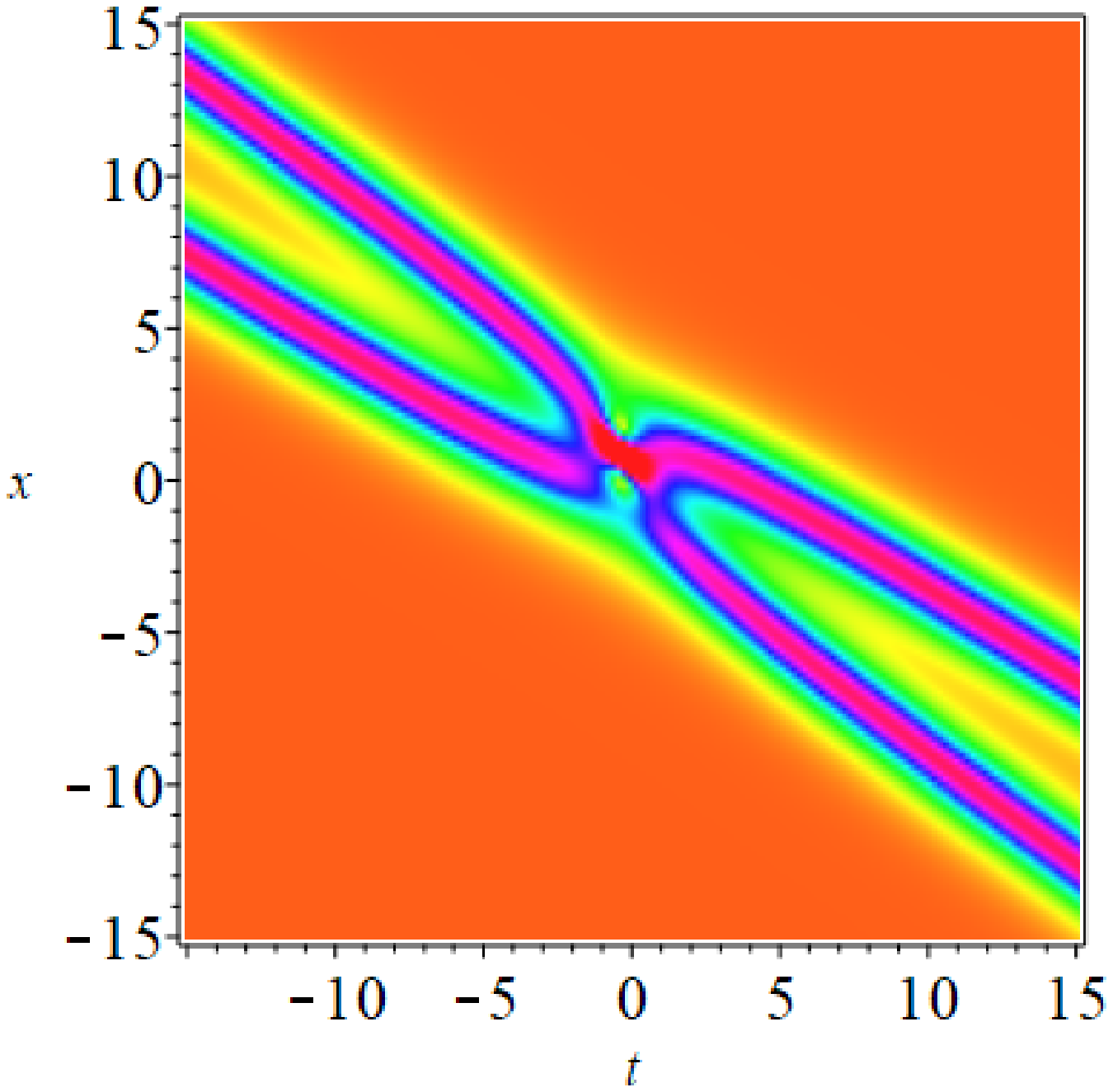}}}
\qquad\quad
{\rotatebox{0}{\includegraphics[width=3.1cm,height=3.1cm,angle=0]{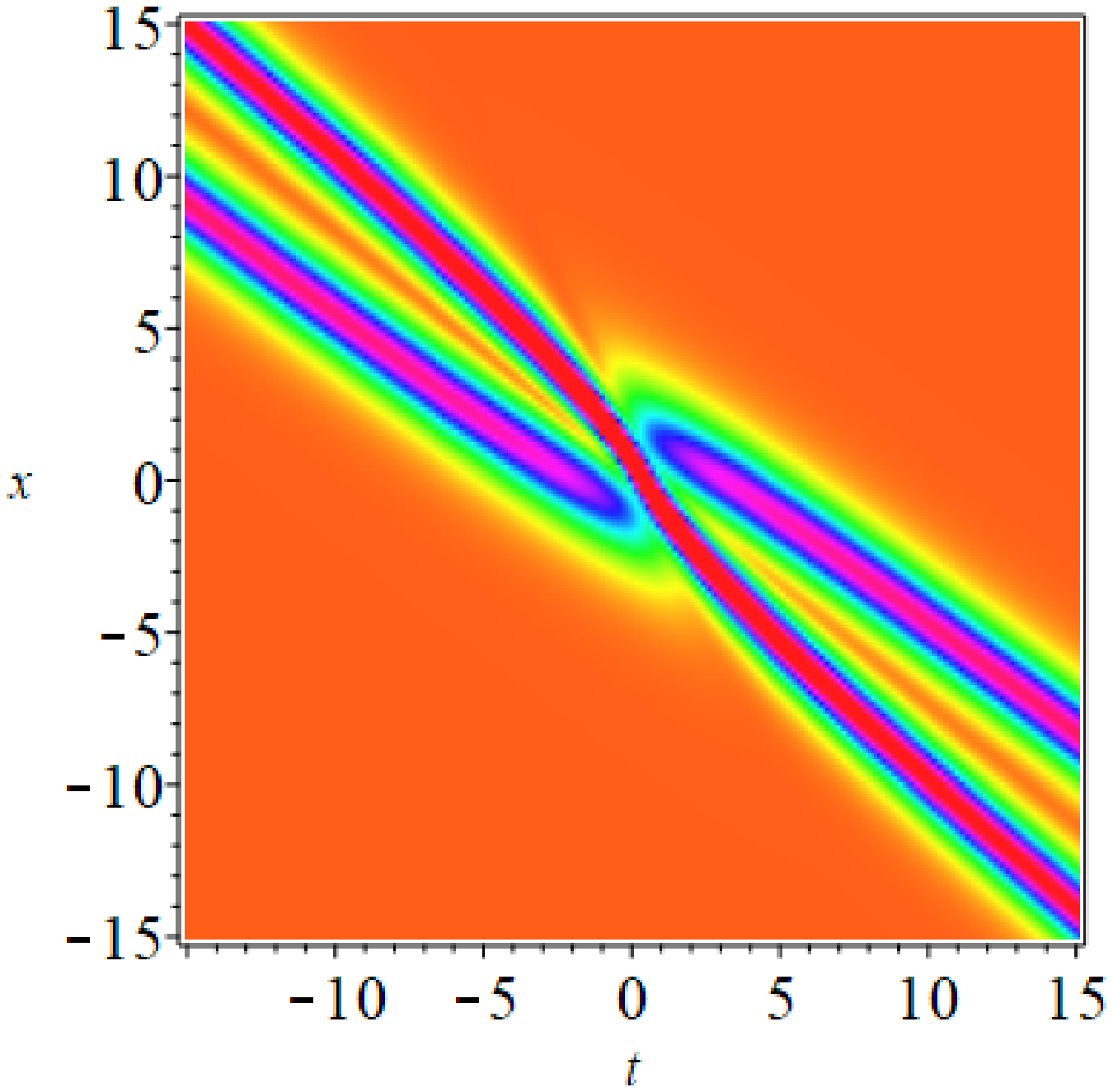}}}
\qquad\quad
{\rotatebox{0}{\includegraphics[width=3.1cm,height=3.1cm,angle=0]{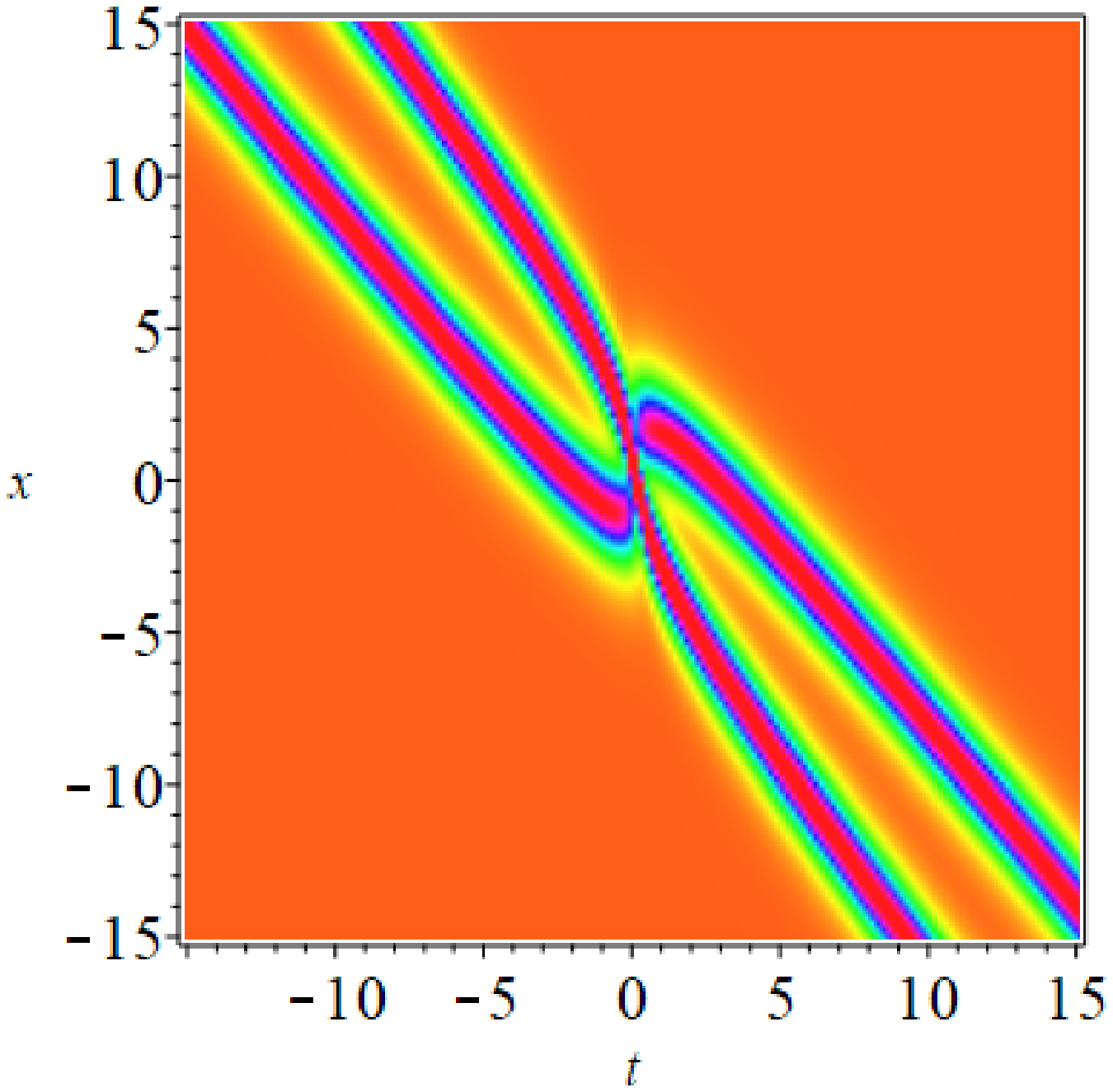}}}
\qquad\quad
{\rotatebox{0}{\includegraphics[width=3.1cm,height=3.1cm,angle=0]{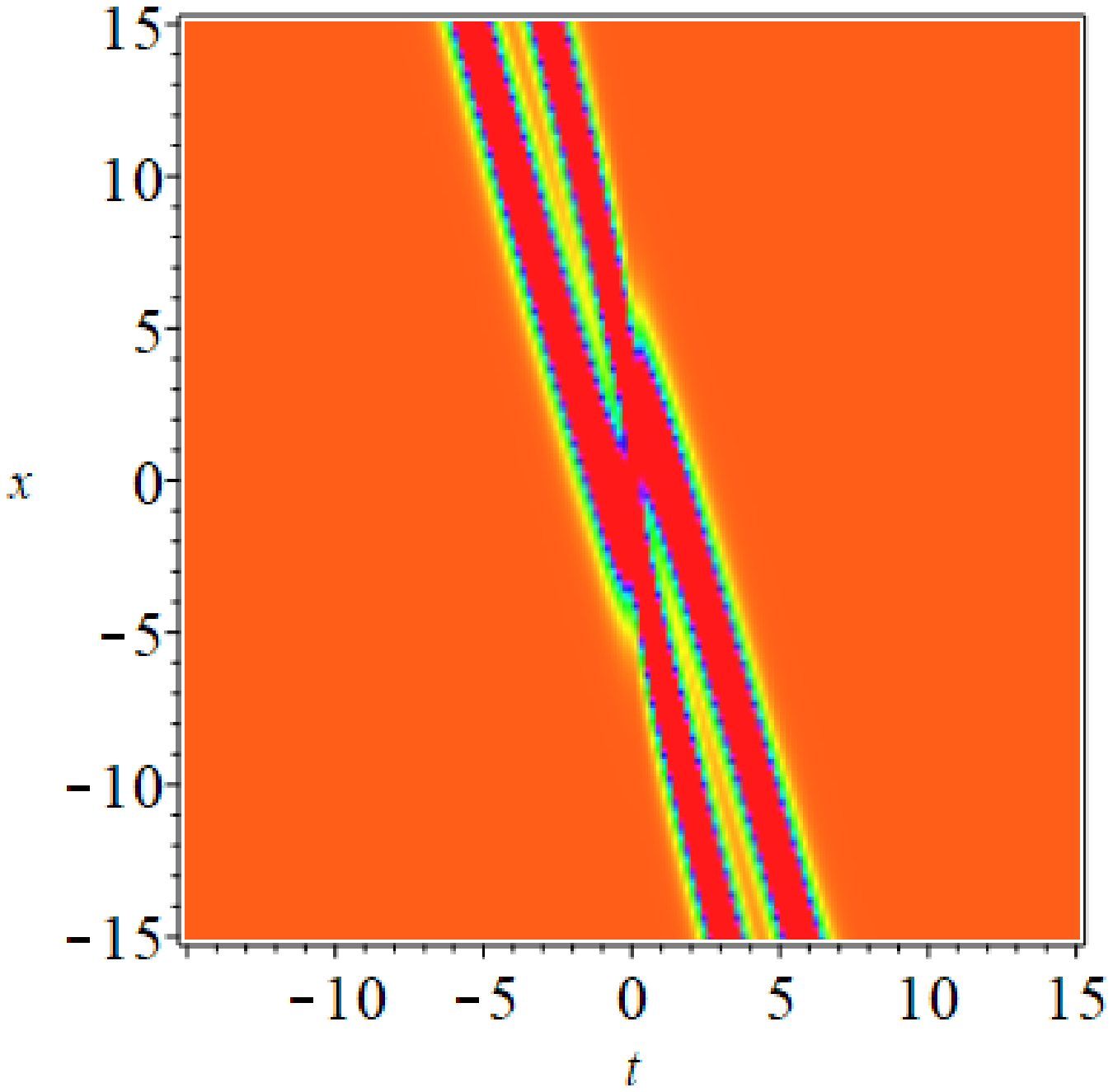}}}\\

\qquad\qquad\qquad$(\textbf{a})\qquad\qquad\qquad\qquad\quad\qquad(\textbf{b})
\qquad\qquad\qquad\qquad\qquad\quad(\textbf{c})\qquad\qquad\qquad\qquad\qquad\quad(\textbf{d})$\\
\noindent { \small \textbf{Figure 2.} (Color online)The density plots of the bound-state soliton solutions \eqref{49a} for Eq.\eqref{1} with the parameters $a=\frac{1}{3}, b=\frac{2}{3}$ and $\textbf{(a)}$ $\delta=0$;
$\textbf{(b)}$ $\delta=\frac{1}{25}$;
$\textbf{(c)}$ $\delta=\frac{1}{5}$;
$\textbf{(d)}$ $\delta=1$.}\\

In fact, the bound-state soliton solutions \eqref{49a} represent the
interaction of two-soliton solutions, and the high peak
comes from the interaction of two solitons under the degeneration of the associated eigenvalues. The corresponding evolution process  of the solutions at different dispersion coefficient $\delta$ are discussed in Figs. 2. As shown in Fig. 2, we easily find that parameter $\delta$ affects the phase positions of two solitons.

\section{The IST with NZBCs and rogue wave}
In this section, we aim to seek the rogue wave solution $q(x, t)$ for the SONLS Eq.\eqref{1} with the following NZBCs as infinity
\begin{align}\label{3}
\lim_{x\rightarrow \pm\infty}q(x, t)=Ae^{i(\alpha x+\beta t)},
\end{align}
where $\alpha$ and $A>0$ are all real constants, $\beta=A^{2}-\frac{1}{2}\alpha^{2}-\delta(\alpha^{6}+90A^{4}\alpha^{2}-30A^{2}\alpha^{4}-20A^{6})$.
\subsection{The construction of the RHP with NZBCs}
We devote  to derive the RHP of the SONLS equation\eqref{1} with NZBCs via the robust inverse scattering transform.  For any time $t$, as $x\rightarrow \pm\infty$, the $q(x, t)$ would tend to a plane wave $Ae^{i(\alpha x+\beta t)}$.
According to the gauge transformation $\phi(x,t)=e^{-i(\alpha x+\beta t)\sigma_{3}/2}\psi(x,t)$, the Lax pair \eqref{3.1} can change into
\begin{align}\label{6}
\psi_{x}=X\psi,\qquad \psi_{t}=T\psi,
\end{align}
where
\begin{gather}
X=i(k+\frac{\alpha}{2})\sigma_{3}+Q_{1},\qquad Q_{1}=\left(\begin{array}{cc}
    0  &  iq^{\ast}e^{i(\alpha x+\beta t)}\\
    iqe^{-i(\alpha x+\beta t)} &  0\\
\end{array}\right),\notag\\
T=e^{i(\alpha x+\beta t)\sigma_{3}/2}(V+\frac{i\beta\sigma_{3}}{2})e^{-i(\alpha x+\beta t)\sigma_{3}/2}.\label{7}
\end{gather}
Let $x\rightarrow\pm\infty$, and using the NZBCs\eqref{3}, the following Lax pair will be satisfied
\begin{align}\label{8}
\psi_{\pm, x}=X_{\pm}\psi_{\pm},\qquad \psi_{\pm t}=T_{\pm}\psi_{\pm},
\end{align}
where
\begin{gather}
X_{\pm}=i(k+\frac{\alpha}{2})\sigma_{3}+Q_{0},\quad Q_{0}=\left(\begin{array}{cc}
    0  &  iA\\
    iA &  0\\
\end{array}\right),\notag\\
 T_{\pm}=\omega(k)X_{\pm}=\left[32\delta k^{5}-16\delta\alpha k^{4}-(16\delta A^{2}-8\delta\alpha^{2})k^{3}+(24\delta\alpha A^{2}-4\delta\alpha^{3})k^{2}\right.\notag\\
 \left.+(1-24\delta\alpha^{2}A^{2}+2\delta\alpha^{4}+12\delta A^{4})k+20\delta\alpha^{3}A^{2}-30\delta\alpha A^{4}-\delta\alpha^{5}-\frac{\alpha}{2}\right]X_{\pm}.
\label{9}
\end{gather}
The fundamental solution of this lax pair is
\begin{align}\label{10}
\psi_{\pm}(k; x, t)=n(k)\left(\begin{array}{cc}
    1  &  \frac{k+\frac{\alpha}{2}-\rho}{A}\\
    -\frac{k+\frac{\alpha}{2}-\rho}{A} &  1\\
\end{array}\right)e^{i\rho(k) \theta(k; x, t)\sigma_{3}}=Y(k)e^{i\rho(k) \theta(k; x, t)\sigma_{3}},
\end{align}
where
\begin{align}\label{11}
\rho(k)^{2}=(k+\frac{\alpha}{2})^{2}+A^{2},\quad n(k)^{2}=\frac{\rho+k+\frac{\alpha}{2}}{2\rho},\quad
\theta(k; x, t)=x+\omega(k)t.
\end{align}
The branch cut of $\rho(k)$
is $\eta=\eta_{+}\cup \eta_{-}$ with $\eta_{+}=[iA-\frac{\alpha}{2}, -\frac{\alpha}{2}]$ and $\eta_{-}=[-iA-\frac{\alpha}{2},-\frac{\alpha}{2}]$, of which the branch cut $\eta$ is oriented upward(see Fig. 3).\\

\centerline{\begin{tikzpicture}[scale=1.1]
\draw[->][thick](-3,0)--(-2,0);
\draw[-][thick](-2.0,0)--(0,0);
\draw[fill] (0.2,-0.2) node{$-\frac{\alpha}{2}$};
\draw[->][thick](0,0)--(2,0);
\draw[-][thick](2,0)--(3,0);
\draw[fill] (0,0) circle [radius=0.035];
\draw [-,very thick] (0,0)--(0,1);
\draw [<-,very thick] (0,1)--(0,2);
\draw [->,very thick] (0,-2)--(0,-1);
\draw [-,very thick] (0,-1)--(0,0);
\draw[fill] (3,0.2) node{$\mathbb{R}$};
\draw[fill] (0.5,2) node{$iA-\frac{\alpha}{2}$};
\draw[fill] (0.5,-2) node{$-iA-\frac{\alpha}{2}$};
\draw[fill] (0.3,1) node{$\eta_{+}$};
\draw[fill] (0.3,-1) node{$\eta_{-}$};
\end{tikzpicture}}
\centerline{\noindent {\small \textbf{Figure 3.} (Color online) The contour $\Sigma_{0}=\mathbb{R}\cup\eta$ of the basic RHP.}}

Then, we assume that $\Psi_{\pm}(x,t,k)$ also solve the Lax pair in Eq.\eqref{6}, and the asymptotic conditions $\Psi_{\pm}(x,t,k)\rightarrow\psi_{\pm}(x,t,k)$ as $x\rightarrow\infty$ should be satisfied. Moreover, taking transformation
\begin{align}\label{12}
\mu_{\pm}(x,t,k)=\Psi_{\pm}(x,t,k)e^{-i\rho\theta\sigma_{3}},
\end{align}
we have
\begin{align}\label{13}
\mu_{\pm}(x,t,k)\rightarrow Y,\qquad x\rightarrow\pm\infty.
\end{align}
Then it is not hard to calculate the following expression for $\mu_{\pm}$, given by
\begin{gather}
\left(Y^{-1}\mu_{\pm}\right)_{x}+i\rho[Y^{-1}\mu_{\pm},\sigma_{3}]
=Y^{-1}(Q_{1}-Q_{0})\mu_{\pm},\notag\\
\left(Y^{-1}\mu_{\pm}\right)_{t}+i\rho\omega[Y^{-1}\mu_{\pm},\sigma_{3}]
=Y^{-1}(T-T_{\pm})\mu_{\pm},\label{14}
\end{gather}
which can be solved by two Volterra integral equations
\begin{align}\label{16}
\mu_{-}(x,t,k)=Y+\int_{-\infty}^{x}Ye^{i\rho(x-y)\hat{\sigma}_{3}}
\left[Y^{-1}(Q_{1}-Q_{0})\mu_{-}(y,t,k)\right]dy,\notag\\
\mu_{+}(x,t,k)=Y-\int_{x}^{+\infty}Ye^{i\rho(x-y)\hat{\sigma}_{3}}
\left[Y^{-1}(Q_{1}-Q_{0})\mu_{+}(y,t,k)\right]dy.
\end{align}

We first construct the RHP through the  robust inverse scattering transform pioneered by Bilman and Miller\cite{Cheng10}.
At the beginning, we suppose $q(x,t)-Ae^{\alpha x+\beta t}\in L^{1}(\mathbb{R})$ and make $\mu_{\pm}=[\mu_{\pm 1},\mu_{\pm 2}]$. Since the first column of $\mu_{-}$ containing the exponential function $e^{-2i\rho(x-y)}$, it can be verified that the
first column of $\mu_{-}$ is analytical on  $\mathbb{C}_{-}\setminus \eta_{-}$,  of which $\mathbb{C}_{-}=\{k: \mbox{Im}(\rho)<0\}$. Analogously, we find that the second column of $\mu_{-}$ is analytical on  $\mathbb{C}_{+}\setminus \eta_{+}$,  where $\mathbb{C}_{+}=\{k: \mbox{Im}(\rho)>0\}$. In conclusion, the $\mu_{-1}$ and $\mu_{+2}$ can be analytically continuous to $\mathbb{C}_{-}\setminus \eta_{-}$, whereas $\mu_{-2}$ and $\mu_{+1}$ can be analytically continuous to $\mathbb{C}_{+}\setminus \eta_{+}$.

Due to $\Psi_{\pm}(x,t,k)$ admit the Lax pair \eqref{6} for $k\in \Sigma_{0}\setminus \left\{-\frac{\alpha}{2}\pm iA\right\}$, the following scattering relation is presented through scattering matrix $S(k)$
\begin{align}\label{17}
\Psi_{+}(x,t,k)=\Psi_{-}(x,t,k)S(k),\qquad k\in \Sigma_{0}\setminus \left\{-\frac{\alpha}{2}\pm iA\right\},
\end{align}
the scattering matrix $S(k)$ can be expressed as
\begin{align}\label{18}
S(k)=\left(\begin{array}{cc}
   S_{11}(k)  &  S_{12}(k)\\
   S_{21}(k) &  S_{22}(k)\\
\end{array}\right),\qquad S_{11}(k)S_{22}(k)-S_{12}(k)S_{21}(k)=1,
\end{align}
where $S_{22}(k)=S_{11}(k^{\ast})^{\ast}, S_{12}(k)=-S_{21}(k^{\ast})^{\ast}$. Then the Beals-Coifman simultaneous solution of \eqref{6} is
\begin{equation}\label{19}
\Phi^{BC}(k; x, t)=\begin{cases}
\left[\frac{\Psi_{+1}(k; x, t)}{S_{11}(k)},\Psi_{-2}(k; x, t)\right],\quad k\in \mathbb{C}_{+}\backslash \eta_{+}\\
\left[\Psi_{-1}(k; x, t),\frac{\Psi_{+2}(k; x, t)}{S_{22}(k)}\right],\quad k\in \mathbb{C}_{-}\backslash \eta_{-}.
\end{cases}
\end{equation}
Let $M^{BC}(k; x, t)=\Phi^{BC}(k; x, t)e^{-i\rho\theta\sigma_{3}}$, the jumping curve for $M^{BC}(k; x, t)$ is the $\mathbb{R}\cup\eta$. Following the similar computations shown in\cite{ZhouX1, ZhouX2}, we can construct another simultaneous solution of the lax pair \eqref{6} for smaller $k$(let's make it $\epsilon$ here) to make this solution no singularities, given by
\begin{equation}\label{20}
\Phi(k; x, t)=\begin{cases}
\Phi^{BC}(k; x, t),\quad k\in D_{+}\cup D_{-},\\
\Phi^{in}(k; x, t),\quad k\in D_{0},
\end{cases}
\end{equation}
where $\Phi^{BC}(k; x, t)$ represents the Beals-Coifman simultaneous solution. $\Phi^{in}(k; x, t)$ is an entire function and can be redefined as $\Phi(k; x, t)\Phi(k; L, 0)^{-1}$. $D_{0}$ on behalf of the open disk whose boundary is $\Sigma_{+}\cup\Sigma_{-}$ and radius is $\epsilon$. Noteworthily, we should  select the  appropriate $\epsilon$ to
make  the scattering data $S_{11}(k), S_{22}(k)$ not be zero on the outside of the disk. Simultaneously, the branch cut $\eta$ should be contained in this disk. As well as, the related domains $D_{\pm}=\left\{k\in \mathbb{C}:|k|\geq \epsilon, \mbox{Im}(k)\gtrless 0\right\}$ and $\Sigma=(-\infty, -\epsilon]\cup [\epsilon, +\infty)\cup\Sigma_{+}\cup\Sigma_{-}$ are shown(see Fig. 4). Set $M(k; x, t)=\Phi(k; x, t)e^{-i\rho\theta\sigma_{3}}$, then the RHP of the SONLS equation with NZBCs is:

\noindent \textbf{Riemann-Hilbert Problem 2}  \emph{
$M(k; x, t)$ solve the following RH problem:
\begin{align}\label{21}
\left\{
\begin{array}{lr}
M(k; x, t)\ \mbox{is analytic in} \ \mathbb{C}\setminus \left\{\Sigma\cup \eta\right\},\\
M_{+}(k; x, t)=\left\{
\begin{array}{lr}M_{-}(k; x, t)e^{i\rho\theta\sigma_{3}}J(k; x, t)e^{-i\rho\theta\sigma_{3}}, \qquad k\in\Sigma,\\
M_{-}(k; x, t)e^{2i\rho_{-}\theta\sigma_{3}},  \qquad k\in\eta,
  \end{array}
\right. \\
M(k; x, t)\rightarrow \mathbb{I},\qquad k\rightarrow \infty,
  \end{array}
\right.
\end{align}
of which the jump matrix $J(k; x, t)$ given in what follows
\begin{align}\label{22}
J(k; x, t)=\left\{
\begin{array}{lr}
\left[\frac{\Psi_{+1}(k; L, 0)}{S_{11}(k)},\Psi_{-2}(k; L, 0)\right],\quad k\in\Sigma_{+}\\
\left[\Psi_{-1}(k; L, 0),\frac{\Psi_{+2}(k; L, 0)}{S_{22}(k)}\right]^{-1},\quad k\in\Sigma_{-} \\
\left[\begin{array}{cc}
   1+|R(k)|^{2}  &  R^{\ast}(k)\\
    R(k) &  1\\
\end{array}\right],\quad k\in(-\infty, -\epsilon]\cup [\epsilon, +\infty),
  \end{array}
\right.
\end{align}
where $R(k)=\frac{S_{21}(k)}{S_{11}(k)}$ and the contour is shown in Fig. 4. Then one can also recover the solution of the Eq.(1) in the form
\begin{align}\label{23}
q(x, t)=2\lim_{k\rightarrow\infty}kM_{21}(k; x, t)e^{i(\alpha x+\beta t)}.
\end{align}
}
\centerline{\begin{tikzpicture}[scale=1.5]
\draw [fill][blue] [-,very thick] (2,0)--(2,0.2);
\draw [fill][blue] [<-,very thick] (2,0.2)--(2,0.5);
\draw [fill][blue] [->,very thick] (2,-0.5)--(2,-0.2);
\draw [fill][blue] [-,very thick] (2,-0.2)--(2,0);
\draw[fill] (2.5,0.5) node{$-\frac{\alpha}{2}+iA$};
\draw[fill] (2.5,-0.5) node{$-\frac{\alpha}{2}-iA$};
\draw[fill] (2.2,0.2) node{$\eta_{+}$};
\draw[fill] (2.2,-0.2) node{$\eta_{-}$};
\draw[->][thick](0.5,0)--(1,0);
\draw[-][thick](1,0)--(1.5,0);
\draw[dashed](1.5,0)--(2.5,0);
\draw[dashed](2.5,0)--(3.5,0);
\draw[->][thick](3.5,0)--(4,0);
\draw[-][thick](4,0)--(4.5,0)node[above]{$\mbox{Re}k$};
\draw[dashed](2.5,2)node[right]{$\mbox{Im}k$}--(2.5,0);
\draw[dashed](2.5,0)--(2.5,-2);
\draw[fill] (1.3,0.05) node[below]{$-\epsilon$};
\draw[fill] (3.6,0.05) node[below]{$\epsilon$};
\draw[fill][red] (2.6,-0.1) node[right]{$D_{0}$};
\draw[fill] (2.5,1.2) node[right]{$\Sigma_{+}$};
\draw[fill] (2.5,-1.2) node[right]{$\Sigma_{-}$};
\draw[fill][red] (3.6,1.3) circle [radius=0] node[right]{$D_{+}$};
\draw[fill] [red] (3.6,-1.3) circle [radius=0] node[right]{$D_{-}$};
\draw[-][thick](3.5,0) arc(0:360:1);
\draw[-<][thick](3.5,0) arc(0:30:1);
\draw[-<][thick](3.5,0) arc(0:150:1);
\draw[->][thick](3.5,0) arc(0:210:1);
\draw[->][thick](3.5,0) arc(0:330:1);
\end{tikzpicture}}
\centerline{\noindent {\small \textbf{Figure 4.} (Color online) Definitions of the regions $D_{0}, D_{\pm}$ and contours $\Sigma, \eta$.}}

\subsection{Rogue waves of SONLS equation}
In this section, we are aimed at the high-order rogue waves of the SONLS equation by using the modified Darboux transformation for the RHP \textbf{2}. Take a gauge transformation in what follows
\begin{align}\label{24}
\widetilde{\Phi}(k; x, t)=\left\{
\begin{array}{lr}
\textbf{G}(k; x, t)\Phi(k; x, t),\quad k\in D_{+}\cup D_{-},\\
\textbf{G}(k; x, t)\Phi(k; x, t)\textbf{G}(k; L, 0)^{-1},\quad k\in D_{0},\\
  \end{array}
\right.
\end{align}
where $\Phi(k; x, t)$ satisfies the Lax pair \eqref{6} and yields $\Phi(k; L, 0)=\mathbb{I}$ for $k\in D_{0}$. The $\textbf{G}$ is defined as
\begin{align}\label{25}
\textbf{G}(k; x, t)=\mathbb{I}+\frac{\textbf{H}(x, t)}{k-\xi}+\frac{\sigma_{2}\textbf{H}^{\ast}(x, t)\sigma_{2}}{k-\xi^{\ast}},
\end{align}
for any point $\xi\in D_{0}$ with $\textbf{H}(x, t)$ being written as
\begin{align}\label{26}
\textbf{H}(x, t)=\frac{-(\xi-\xi^{\ast})^{2}(1-\vartheta^{\ast}(x, t))\textbf{s}(x, t)\textbf{s}^{T}(x, t)\sigma_{2}+(\xi-\xi^{\ast})N(x, t)\sigma_{2}\textbf{s}^{\ast}(x, t)\textbf{s}^{T}(x, t)\sigma_{2}}{-(\xi-\xi^{\ast})^{2}|1-\vartheta(x, t)|^{2}+N^{2}(x, t)},
\end{align}
of which  $\textbf{s}(x, t)=\Phi(\xi; x, t)\textbf{c}$, $N(x, t)=
\textbf{s}^{\dag}(x, t)\textbf{s}(x, t), \vartheta(x, t)=\textbf{s}^{T}(x, t)\sigma_{2}\textbf{s}'(x, t)$, and $\textbf{c}=(c_{1},c_{2})^{T}$  is an
arbitrary column vector. Then, the corresponding jump condition of  matrix $\widetilde{M}(k; x, t)=\widetilde{\Phi}(k; x, t)e^{-i\rho\theta\sigma_{3}}$ make a difference when $k\in\Sigma_{+}\cup\Sigma_{-}$, and the corresponding jump matrix
$J(k)$ is replaced by
\begin{align}\label{27}
\widetilde{J}(k; x, t)=\left\{
\begin{array}{lr}
\textbf{G}(k; L, 0)J(k; x, t),\quad k\in \Sigma_{+},\\
J(k; x, t)\textbf{G}(k; L, 0)^{-1},\quad k\in \Sigma_{-}.\\
  \end{array}
\right.
\end{align}
Similarly, the potential function $\widetilde{q}(x, t)$ can be recovered from the new RHP
$\widetilde{M}(k; x, t)$, that is
\begin{align}\label{28}
\widetilde{q}(x, t)=2\lim_{k\rightarrow\infty}k\widetilde{M}_{21}(k; x, t)e^{i(\alpha x+\beta t)}=q(x, t)+2(\textbf{H}_{21}-\textbf{H}_{12}^{\ast})e^{i(\alpha x+\beta t)}.
\end{align}
In  addition, as we make $\textbf{c}$ become the form $\epsilon^{-1}\textbf{c}_{\infty}$ with
$\textbf{c}_{\infty}\in \mathbb{C}^{2}\setminus\left\{0\right\}$ being a fixed vector and  let $\epsilon$ tend to $0$, then the matrix $\textbf{G}(k; x, t)$ can also be taken a limit process, given by
\begin{align}\label{28.1}
\textbf{G}_{\infty}(k; x, t)=\mathbb{I}+\frac{\textbf{H}_{\infty}(x, t)}{k-\xi}+\frac{\sigma_{2}\textbf{H}_{\infty}^{\ast}(x, t)\sigma_{2}}{k-\xi^{\ast}},
\end{align}
where $\textbf{H}_{\infty}(x, t)=\lim_ {\epsilon\rightarrow\infty} \textbf{H}(x, t)$.

In order to apply the Darboux transformation, the vector $\textbf{s}(x, t)$ should be given first. Therefore, in terms of the background eigenvector
matrix $\Phi_{bg}(k; x, t)=\psi_{\pm}(k; x, t)$, $\Phi_{bg}^{in}(k; x, t)=
\Phi_{bg}(k; x, t)\Phi_{bg}(k; 0, 0)^{-1}$ can be regarded as the basic solutions, then one has
\begin{align}\label{29}
\Phi_{bg}^{in}(k; x, t)=\frac{\sin\left(\rho(k)\theta(k; x, t)\right)}{\rho(k)}X_{\pm}+\cos\left(\rho(k)\theta(k; x, t)\right)\mathbb{I}.
\end{align}
Furthermore, we have
\begin{gather}
\textbf{s}(\xi; x, t)=\Phi_{bg}^{in}(\xi; x, t)\textbf{c}
=i\gamma(\xi, x, t)\left[(\xi+\frac{\alpha}{2})\sigma_{3}\textbf{c}+A\sigma_{1}\textbf{c}\right]+\chi(\xi, x, t)\textbf{c}.\label{30}
\end{gather}
where
\begin{align}\label{31}
\sigma_{1}=\left(\begin{array}{cc}
   0  &  1\\
    1 &  0\\
\end{array}\right),\quad \gamma(\xi, x, t)=\frac{\sin\left(\rho(\xi)\theta(\xi; x, t)\right)}{\rho(\xi)},\quad \chi(\xi, x, t)=\cos\left(\rho(\xi)\theta(\xi; x, t)\right),
\end{align}
and
\begin{gather}
N(x, t)=\textbf{s}^{\dag}(x, t)\textbf{s}(x, t)=\left[(\xi+\frac{\alpha}{2})(\xi^{\ast}+\frac{\alpha}{2})|\gamma|^{2}+|\chi|^{2}+A^{2}|\gamma|^{2}\right]\textbf{c}^{\dag}\textbf{c}\notag\\
+\left[i(\xi+\frac{\alpha}{2})\gamma\chi^{\ast}-i(\xi^{\ast}+\frac{\alpha}{2})\gamma^{\ast}\chi\right]\textbf{c}^{\dag}\sigma_{3}\textbf{c}
+\left[iA\gamma\chi^{\ast}-iA\gamma^{\ast}\chi\right]\textbf{c}^{\dag}\sigma_{1}\textbf{c}
+i(\xi^{\ast}-\xi)A|\gamma|^{2}\textbf{c}^{\dag}\sigma_{2}\textbf{c},\label{32}
\end{gather}
\begin{gather}
\vartheta(x, t)=\textbf{s}^{T}(x, t)\sigma_{2}\textbf{s}'(x, t)
=\left[\gamma\chi'(\xi+\frac{\alpha}{2})-\gamma'\chi(\xi+\frac{\alpha}{2})-\gamma\chi\right]\textbf{c}^{T}\sigma_{1}\textbf{c}\notag\\
+\left[\gamma\gamma'(\xi+\frac{\alpha}{2})^{2}+\gamma^{2}(\xi+\frac{\alpha}{2})+\gamma\gamma'A^{2}+\chi\chi'\right]\textbf{c}^{T}\sigma_{2}\textbf{c}
+(A\gamma'\chi-A\gamma\chi')\textbf{c}^{T}\sigma_{3}\textbf{c}-iA\gamma^{2}\textbf{c}^{T}\textbf{c},\label{33}
\end{gather}
Finally, the solutions of Eq.\eqref{1} are
\begin{align}\label{34}
\widetilde{q}(x, t)=\left(A+\frac{2i(\xi-\xi^{\ast})^{2}\left[(1-\vartheta^{\ast})s_{2}^{2}-(1-\vartheta)s_{1}^{\ast 2}\right]+4(\xi-\xi^{\ast})Ns_{1}^{\ast}s_{2}}{(\xi-\xi^{\ast})^{2}|1-\vartheta|^{2}-N^{2}}\right)e^{i(\alpha x+\beta t)},
\end{align}
where $\textbf{s}, N, \vartheta$ are given in \eqref{30}, \eqref{32}, \eqref{33}. Moreover, setting $\textbf{c}=\textbf{c}_{\infty}\epsilon^{-1}$ with $\epsilon\rightarrow 0$, the solution \eqref{34} is changed into
\begin{align}\label{35}
\widetilde{q}_{\infty}(x, t)=\left(A-\frac{2i(\xi-\xi^{\ast})^{2}\left(\vartheta_{\infty}^{\ast}s_{\infty2}^{2}-\vartheta_{\infty}s_{\infty1}^{\ast 2}\right)-4(\xi-\xi^{\ast})N_{\infty}s_{\infty1}^{\ast}s_{\infty2}}{(\xi-\xi^{\ast})^{2}|\vartheta_{\infty}|^{2}-N_{\infty}^{2}}\right)e^{i(\alpha x+\beta t)},
\end{align}
where $\textbf{s}_{\infty}, N_{\infty}, \vartheta_{\infty}$ are given in \eqref{30}, \eqref{32}, \eqref{33} with $\textbf{c}$ substituted by $\textbf{c}_{\infty}$, respectively.

Based on the theorem of spectral analysis, when choosing different $\xi$, the corresponding solution properties
are very different. For the case of $\xi=-\frac{\alpha}{2}+i\lambda A$ with $|\lambda|>1$, it becomes the temporal-spatial periodic breather waves, which can be attested by Fig. 5. But when $|\lambda|<1$, as we can see in  Fig. 6, it turns into the spatial periodic breather waves.

\qquad
{\rotatebox{0}{\includegraphics[width=3.6cm,height=3.0cm,angle=0]{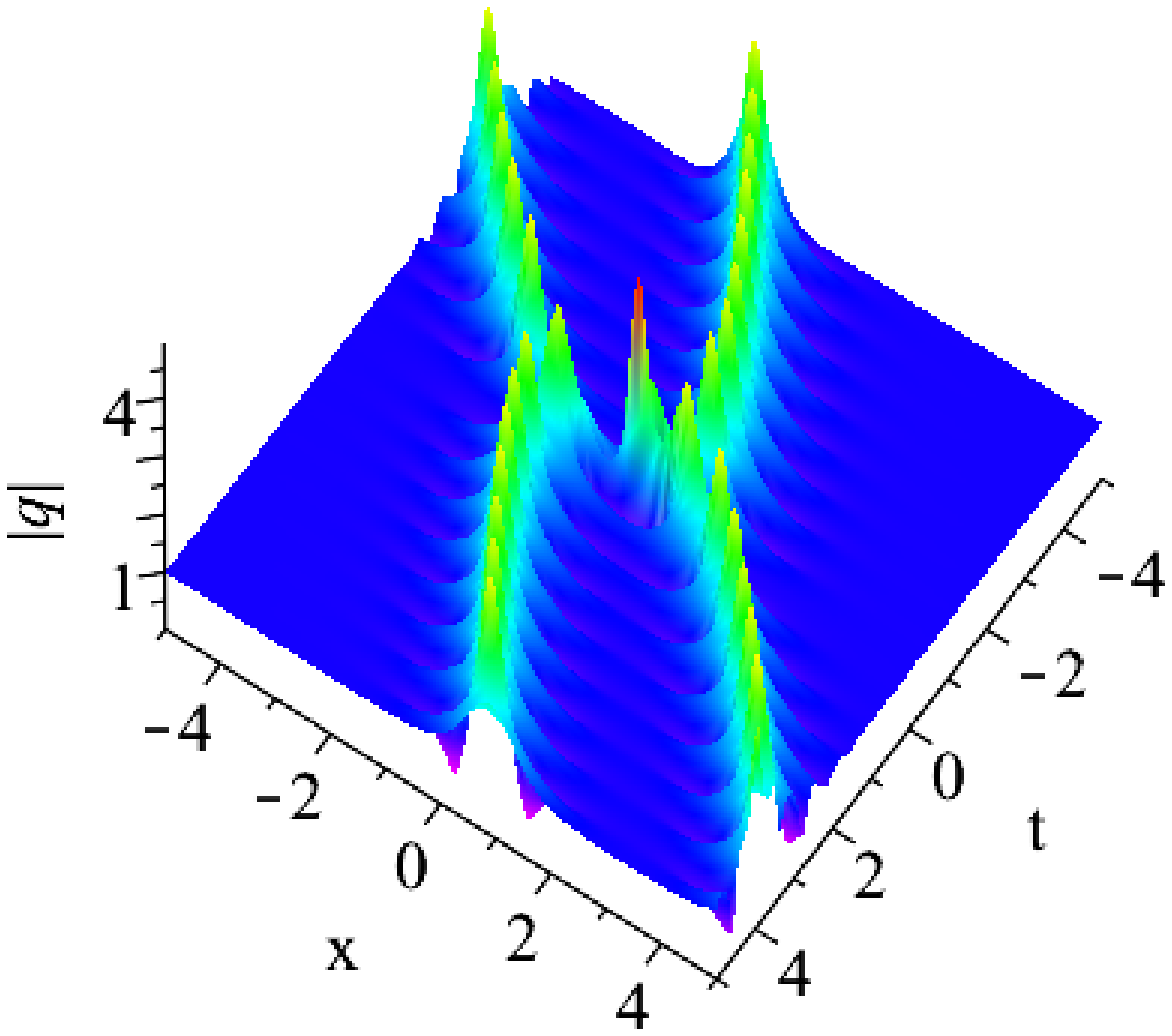}}}
\qquad\qquad\qquad
{\rotatebox{0}{\includegraphics[width=3.6cm,height=3.0cm,angle=0]{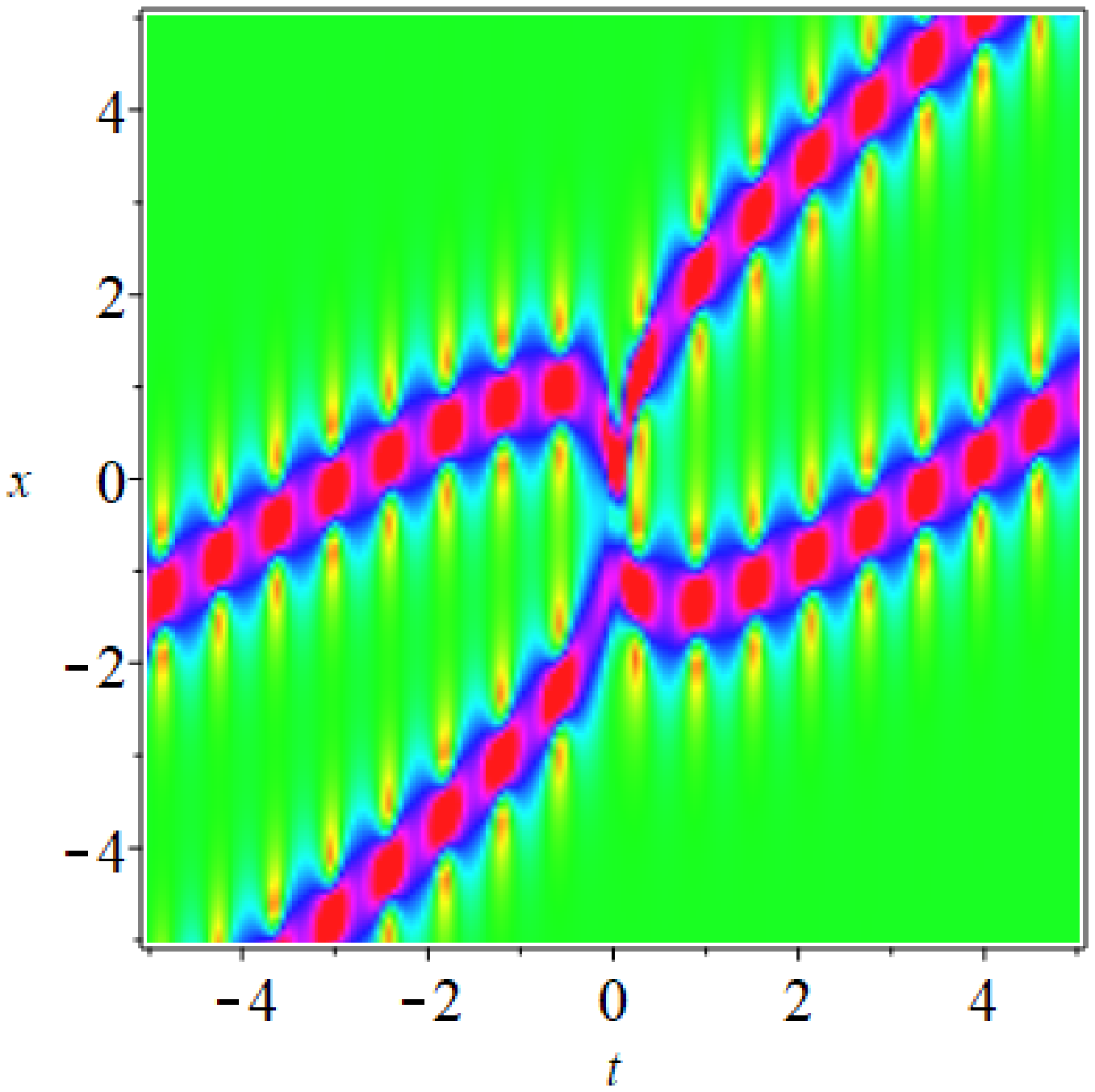}}}
\qquad\qquad\qquad
{\rotatebox{0}{\includegraphics[width=3.6cm,height=3.0cm,angle=0]{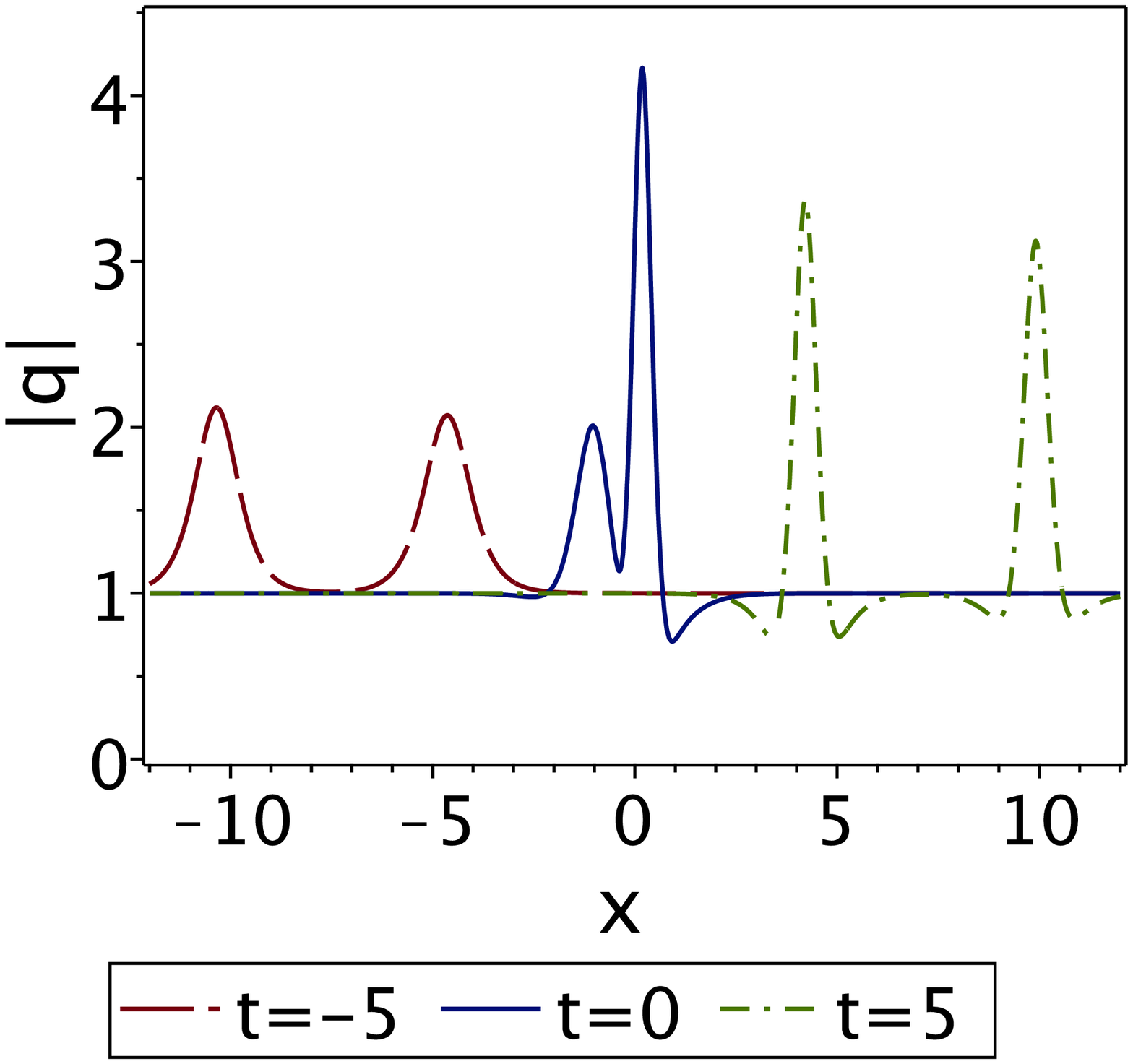}}}\\

\qquad\qquad\qquad$(\textbf{a})\qquad\qquad\qquad\qquad\qquad\qquad\qquad\qquad(\textbf{b})
\qquad\qquad\qquad\qquad\qquad\qquad\qquad\qquad(\textbf{c})$\\
\noindent { \small \textbf{Figure 5.} (Color online) The temporal-spatial periodic breather waves \eqref{34} for Eq.\eqref{1} with the parameters $A=1, \alpha=\frac{1}{10}, \delta=0.01, \lambda=\frac{3}{2}, c_{1}=i, c_{2}=i+1$. $\textbf{(a)}$ Three dimensional plot;
$\textbf{(b)}$ The density plot;
$\textbf{(c)}$ The wave propagation along the $x$-axis with $t =-10$ (long-dashed line), $t = 0$ (solid line), $t = 10$ (dash-dotted line).}\\
%

\qquad
{\rotatebox{0}{\includegraphics[width=3.6cm,height=3.0cm,angle=0]{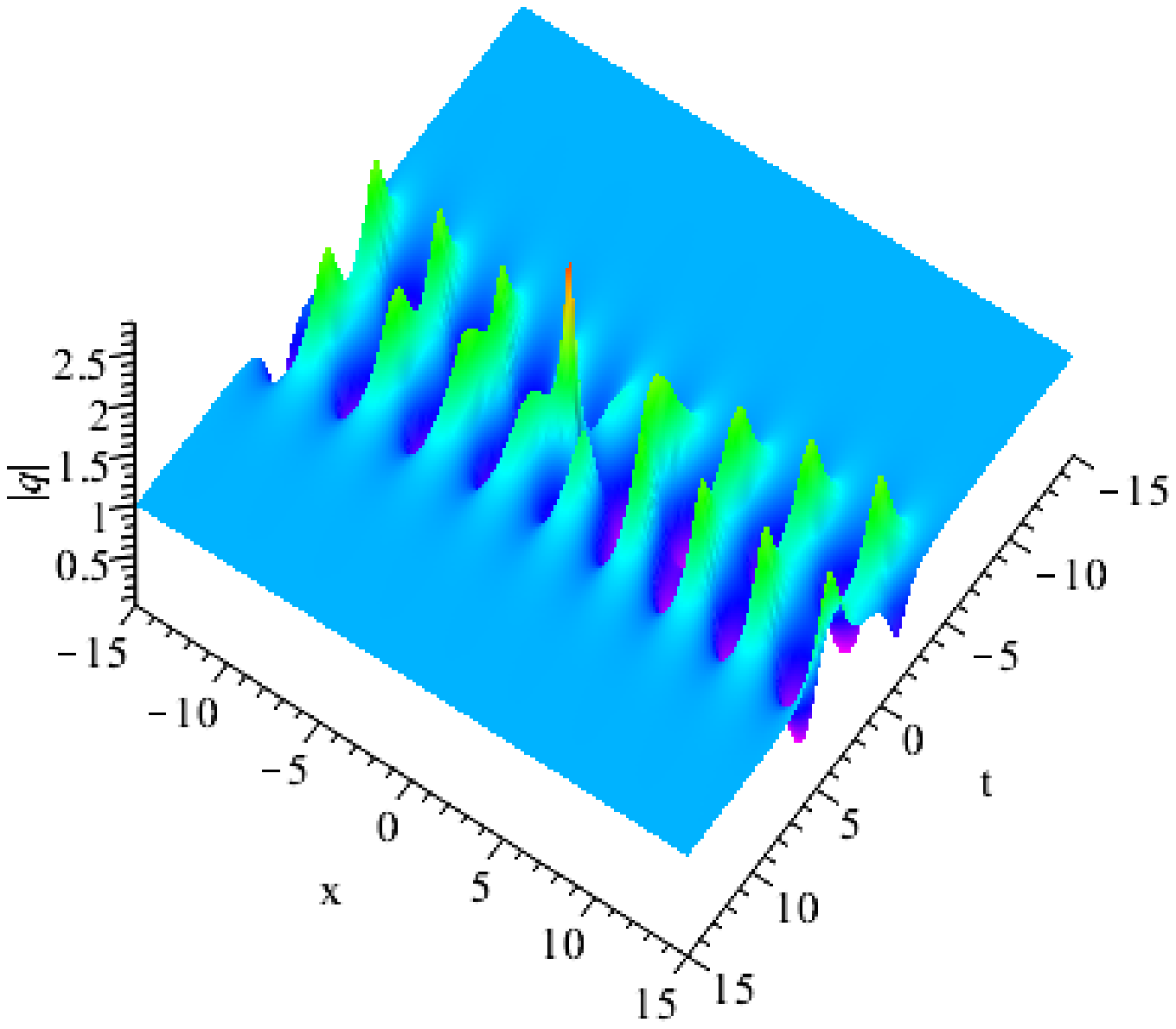}}}
\qquad\qquad\qquad
{\rotatebox{0}{\includegraphics[width=3.6cm,height=3.0cm,angle=0]{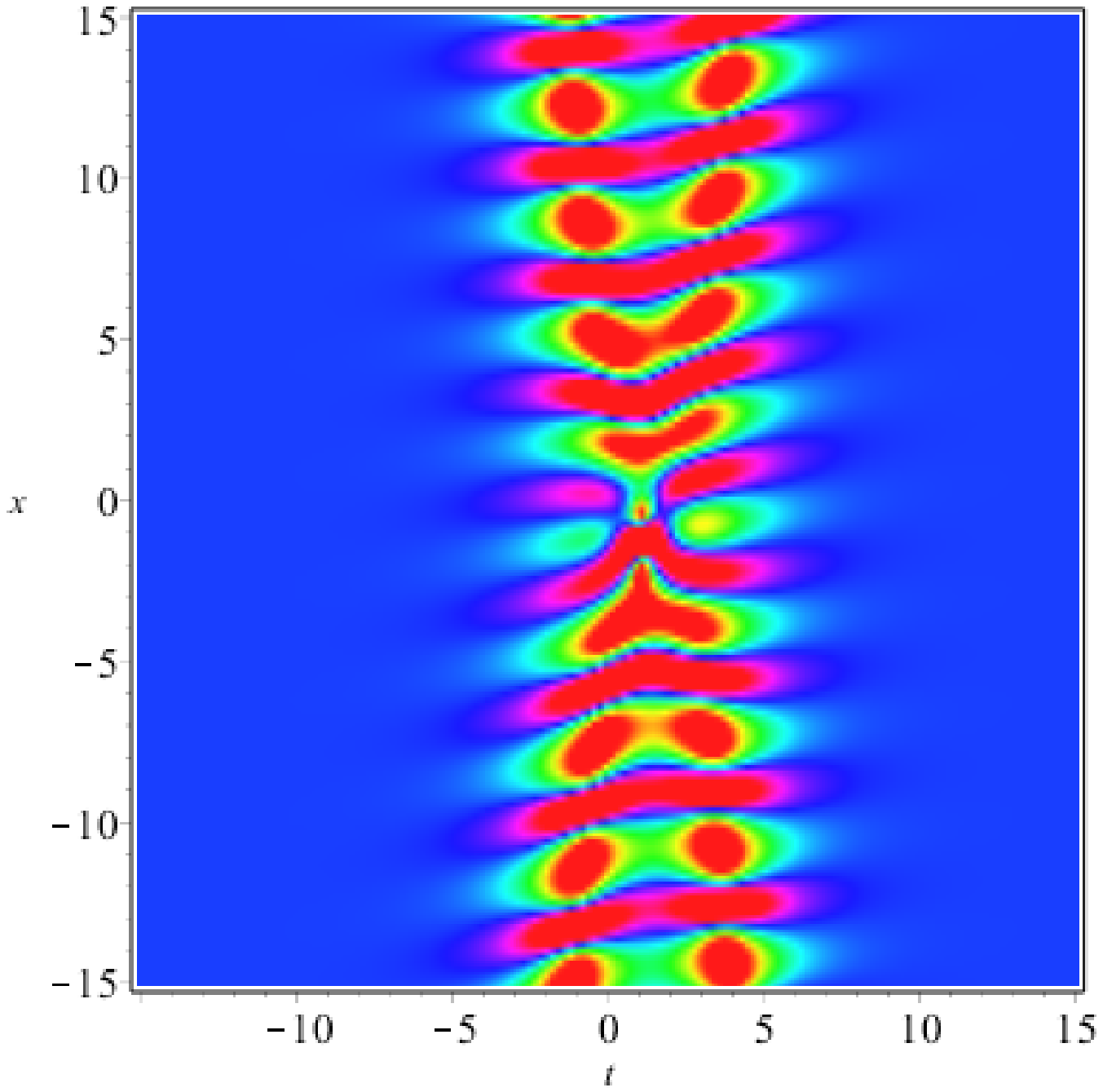}}}
\qquad\qquad\qquad
{\rotatebox{0}{\includegraphics[width=3.6cm,height=3.0cm,angle=0]{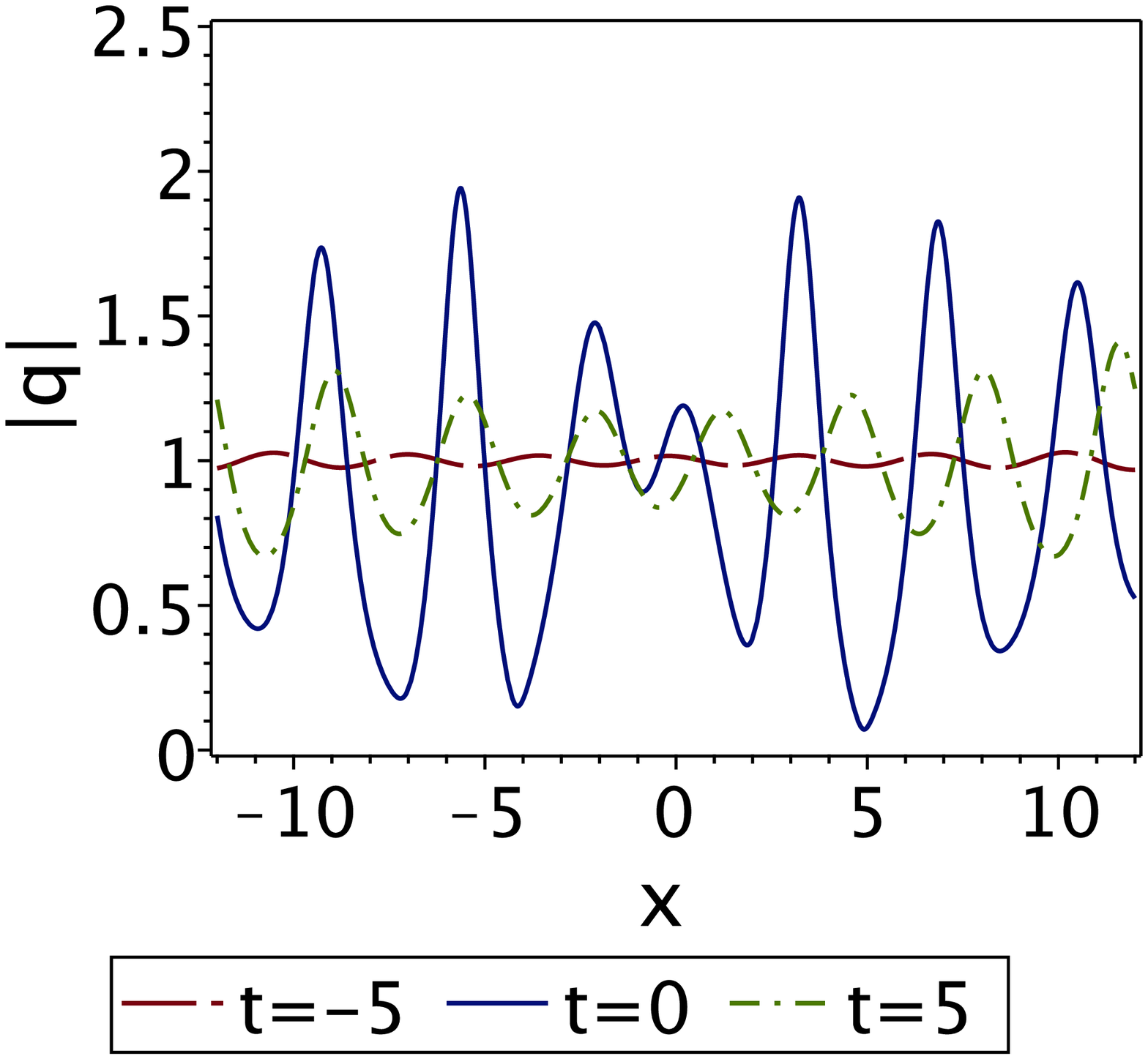}}}\\

\qquad\qquad\qquad$(\textbf{a})\qquad\qquad\qquad\qquad\qquad\qquad\qquad\qquad(\textbf{b})
\qquad\qquad\qquad\qquad\qquad\qquad\qquad\qquad(\textbf{c})$\\
\noindent { \small \textbf{Figure 6.} (Color online) The spatial periodic breather waves \eqref{34} for Eq.\eqref{1} with the parameters $A=1, \delta=0.01, \alpha=\frac{1}{10}, \lambda=\frac{1}{2}, c_{1}=i, c_{2}=i+1$. $\textbf{(a)}$ Three dimensional plot;
$\textbf{(b)}$ The density plot;
$\textbf{(c)}$ The wave propagation along the $x$-axis with $t =-5$ (long-dashed line), $t = 0$ (solid line), $t = 5$ (dash-dotted line).}\\

In order to obtain the rogue waves, we should take $\xi=-\frac{\alpha}{2}\pm iA$(here $|\lambda|=\pm 1$). For the sake of convenience, we study rogue waves with  $\xi=-\frac{\alpha}{2}+iA$(it is similar when $\xi=-\frac{\alpha}{2}-iA$). Then, we have
\begin{align}\label{35.1}
\textbf{s}(x,t)=\left(\begin{array}{c}
   i\theta(\xi)A(ic_{1}+c_{2})+c_{1}\\
   \theta(\xi)A(ic_{1}+c_{2})+c_{2}\\
\end{array}\right),
\end{align}
\begin{align}\label{35.2}
\textbf{s}'(x,t)=\left(\begin{array}{c}
   \frac{1}{3}A(A\theta^{3}+3i\theta')(ic_{1}+c_{2})-ic_{1}\theta(A\theta-1)\\
   -\frac{i}{3}A(A\theta^{3}+3i\theta')(ic_{1}+c_{2})-ic_{2}\theta(A\theta+1)\\
\end{array}\right),
\end{align}
\begin{align}\label{35.3}
N=2A\left(A\theta(c_{1}^{2}+c_{2}^{2})-\frac{1}{2}(c_{1}+ic_{2})^{2}\right)+A\theta(ic_{1}+c_{2})^{2}+c_{1}^{2}+c_{2}^{2},
\end{align}
\begin{align}\label{35.4}
\vartheta=-iA\theta^{2}(c_{1}^{2}+c_{2}^{2})
-A\theta'(ic_{1}+c_{2})^{2}+\frac{2}{3}iA^{2}\theta^{3}(c_{1}^{2}-c_{2}^{2})+(\frac{4}{3}A^{2}\theta^{3}-2\theta)c_{1}c_{2}.
\end{align}
It is not difficult to see that the first-order rogue wave can be derived  when $c_{1}=ic_{2}$. For instance, let $c_{2} = 1$, the first-order rogue wave solution is expressed as
\begin{align}\label{36.1}
\widetilde{q}(x, t)=\left(\frac{2A\left((iA^{2}-2A^{2}\theta-2A)\theta^{\ast}+(2A-iA^{2})\theta-2iA-\frac{A^{2}}{2}+\frac{3}{2}\right)}
{2A^{2}(i-2\theta)\theta^{\ast}-2iA^{2}\theta-A^{2}-1}\right)e^{i(\alpha x+\beta t)}.
\end{align}
In order to make the rogue wave center being at the origin, we set $c_{\infty}=(ic,c)^{T}$ at \eqref{35}. The exact formula of first-order rogue wave can be shown with $c= 1$
\begin{align}\label{36.2}
\widetilde{q}(x, t)=\left(\frac{A(4A^{2}\theta\theta^{\ast}+4A\theta^{\ast}-4A\theta-3)}{4A^{2}\theta\theta^{\ast}+1}\right)e^{i(\alpha x+\beta t)},
\end{align}
whose maximum amplitude is equal to $3A$, which is different with the amplitude given in \eqref{36.1}. Obviously, there are three free parameters $A$ , $\alpha$ and $\delta$ in Eq.\eqref{36.2}. Here, parameter $A$ determines the amplitude of wave and the height of the background wave. Especially, the effects of $\alpha$ and $\delta$ on the dynamic behavior of the rogue wave \eqref{36.2} are analyzed respectively in what follows.

\qquad
{\rotatebox{0}{\includegraphics[width=3.6cm,height=3.0cm,angle=0]{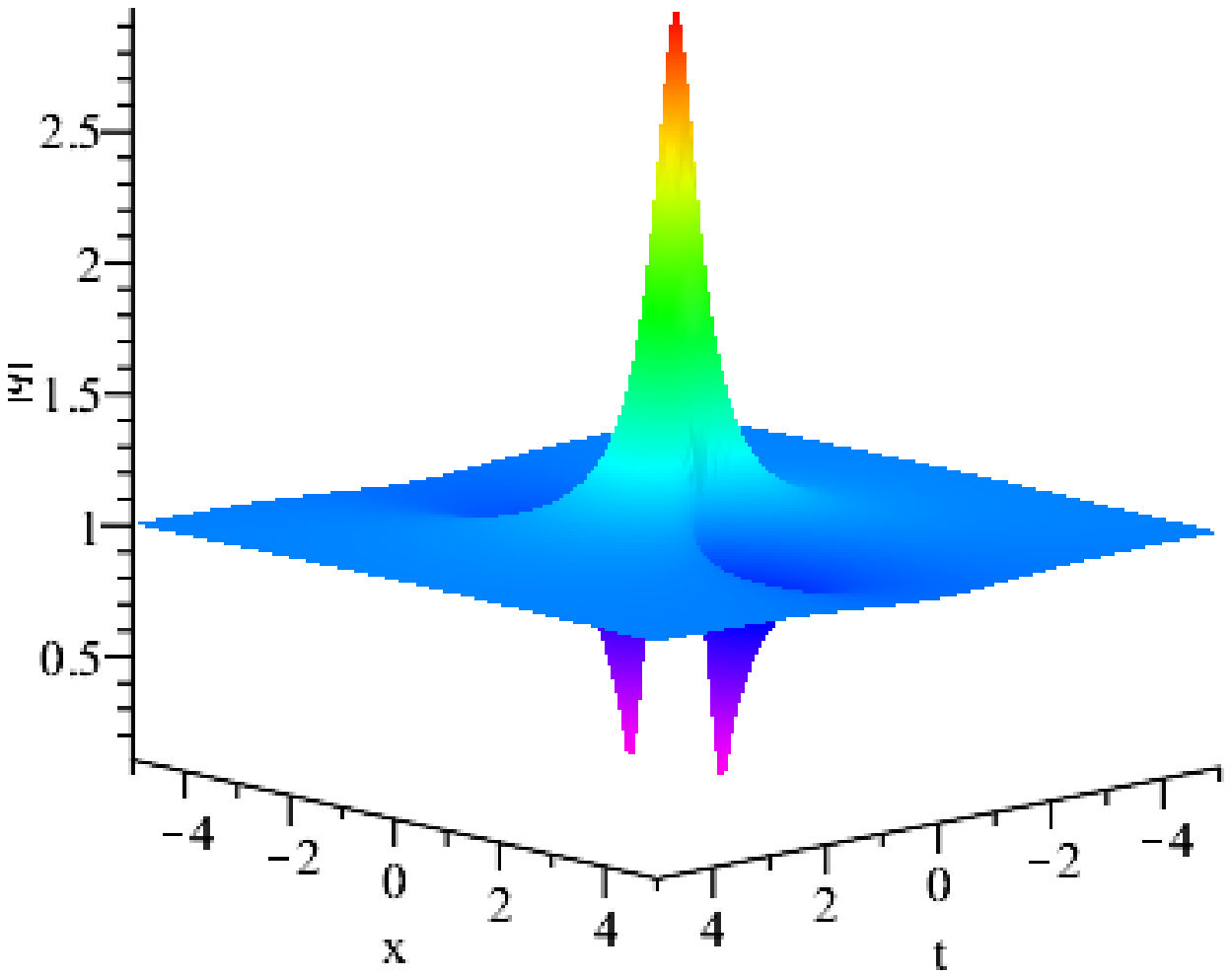}}}
\qquad\qquad\qquad
{\rotatebox{0}{\includegraphics[width=3.6cm,height=3.0cm,angle=0]{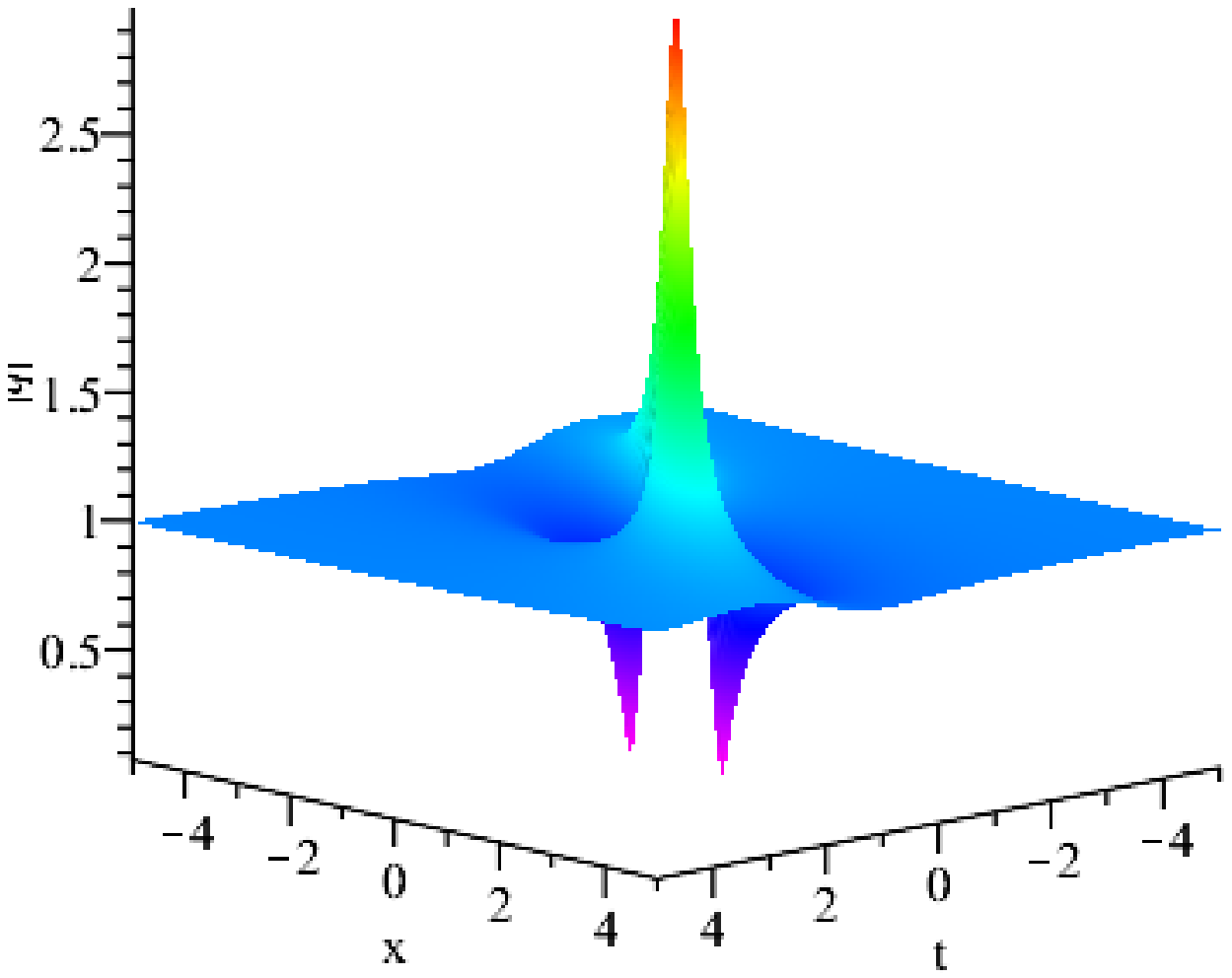}}}
\qquad\qquad\qquad
{\rotatebox{0}{\includegraphics[width=3.6cm,height=3.0cm,angle=0]{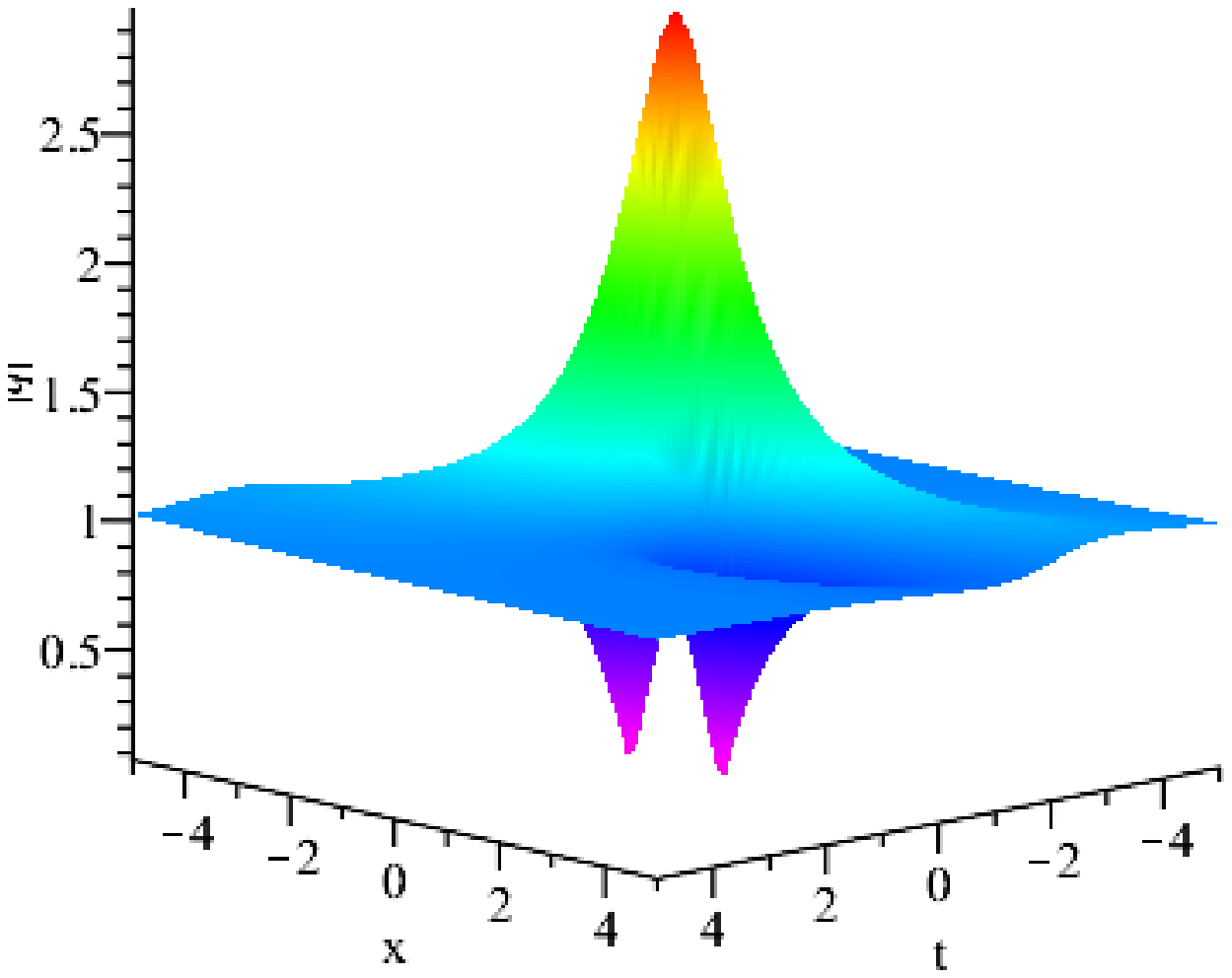}}}\\

\qquad\qquad\qquad$(\textbf{a})\qquad\qquad\qquad\qquad\qquad\qquad\qquad\qquad(\textbf{b})
\qquad\qquad\qquad\qquad\qquad\qquad\qquad\qquad(\textbf{c})$\\

\qquad
{\rotatebox{0}{\includegraphics[width=3.6cm,height=3.0cm,angle=0]{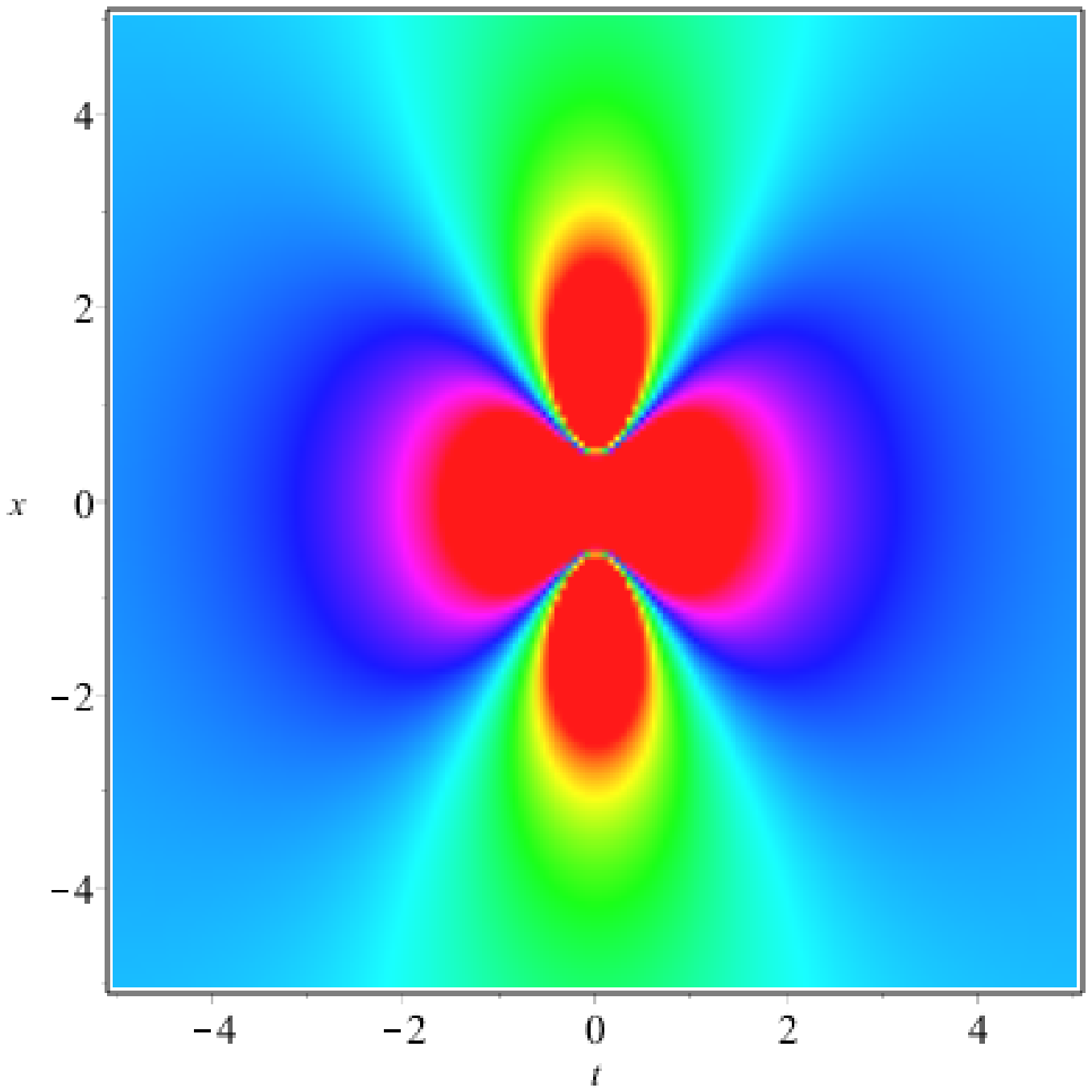}}}
\qquad\qquad\qquad
{\rotatebox{0}{\includegraphics[width=3.6cm,height=3.0cm,angle=0]{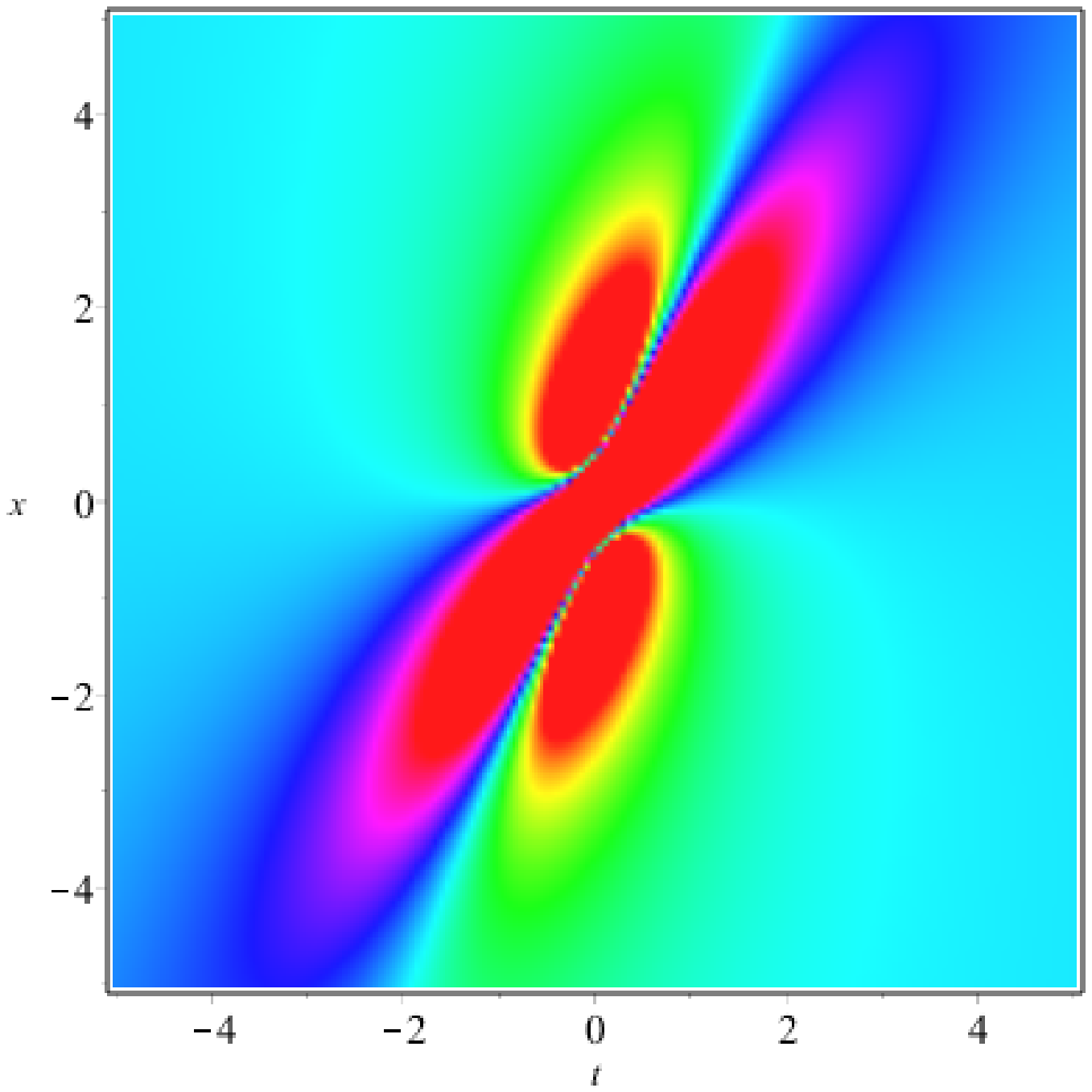}}}
\qquad\qquad\qquad
{\rotatebox{0}{\includegraphics[width=3.6cm,height=3.0cm,angle=0]{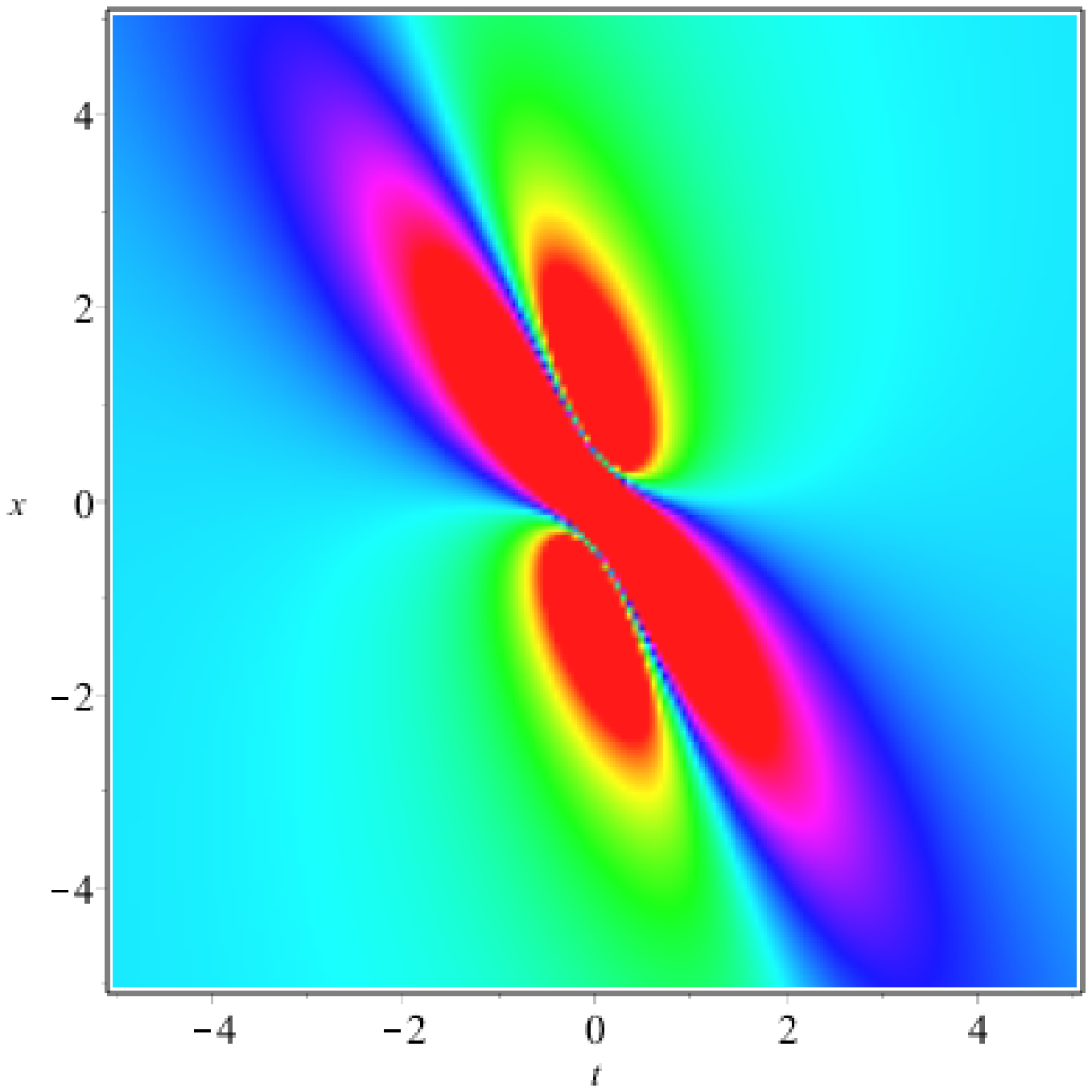}}}\\

\qquad\qquad\qquad$(\textbf{d})\qquad\qquad\qquad\qquad\qquad\qquad\qquad\qquad(\textbf{e})
\qquad\qquad\qquad\qquad\qquad\qquad\qquad\qquad(\textbf{f})$\\
\noindent { \small \textbf{Figure 7.} (Color online) Three dimensional plots and density plots of the first-order rogue wave \eqref{36.2} for Eq.\eqref{1} with the parameters $A=1, \delta=0.01$:
$\textbf{(a,d)}$ $\alpha=0$ ,
$\textbf{(b,e)}$ $\alpha=0.5$,
$\textbf{(c,f)}$ $\alpha=-0.5$.}\\

In Fig.7, fixing $\delta=0.01$, we mainly analyse the effect of $\alpha$ on the wave through different selections of parameter $\alpha$. Compared with  $\alpha=0$, the ridge direction of the rogue waves turns clockwise for $\alpha>0$, and it turns counter clockwise at the case of $\alpha<0$. As well as, the increase of parameter $|\alpha|$ leads to the increase of the angle between the ridge of the rogue waves and the $x$-axis.  As displayed in Fig.8, for fixed parameter $\alpha=0$,  we analyse the corresponding
evolution process of the rogue wave at different dispersion coefficient $\delta$. It is easily to find the fact that the higher order dispersion coefficient $\delta$ affect the phase of rogue waves.  When $\delta$ increases, the ridge of the rogue wave gradually disappears. On the contrary, the length of the trough becomes more and more larger. Besides, the width of the rogue wave decrease as $\delta$ increases. It is worth mentioning that the Fig.8 (a)(d) depict the rogue wave for the standard NLS equation with $\delta=0$.

\qquad
{\rotatebox{0}{\includegraphics[width=3.6cm,height=3.0cm,angle=0]{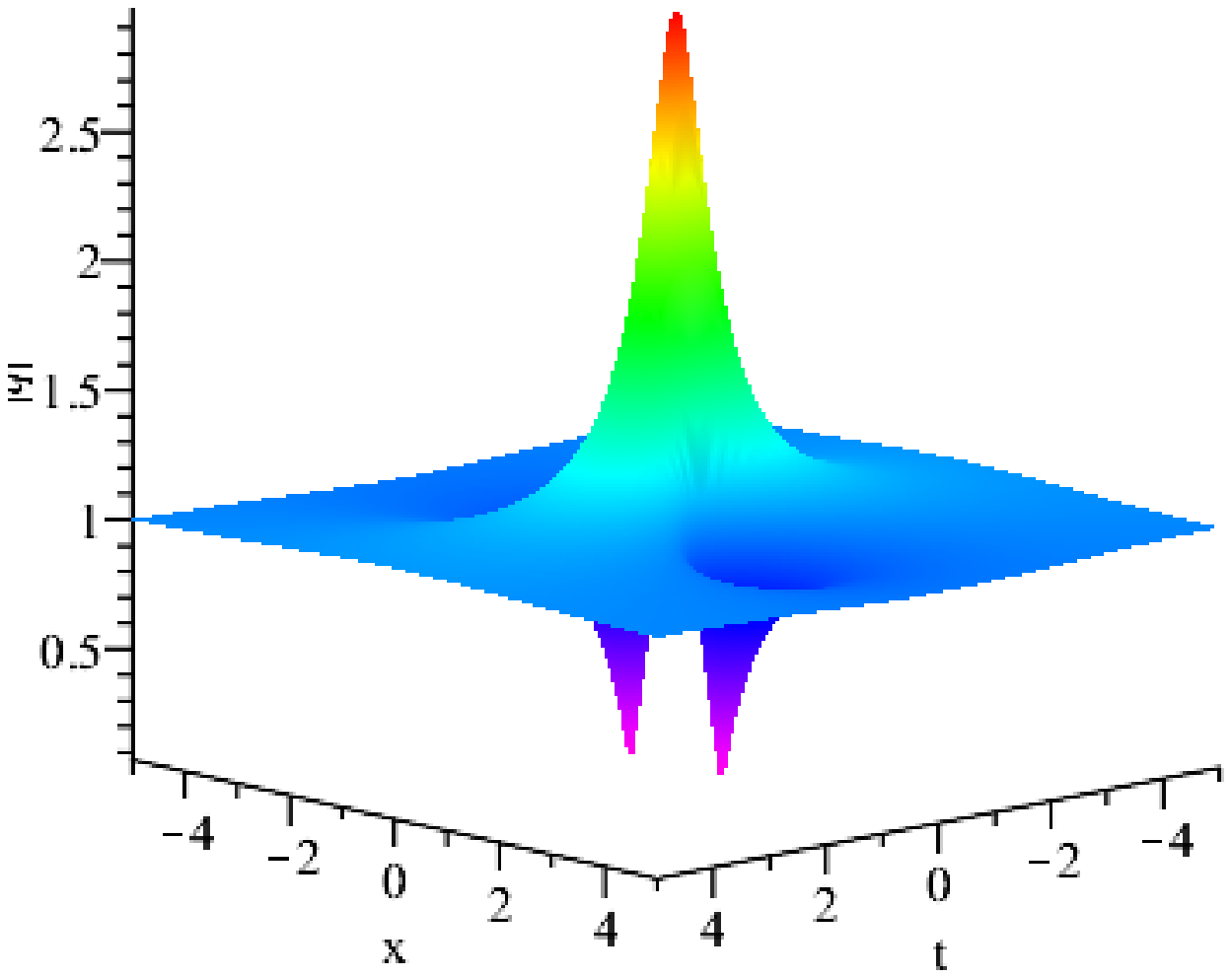}}}
\qquad\qquad\qquad
{\rotatebox{0}{\includegraphics[width=3.6cm,height=3.0cm,angle=0]{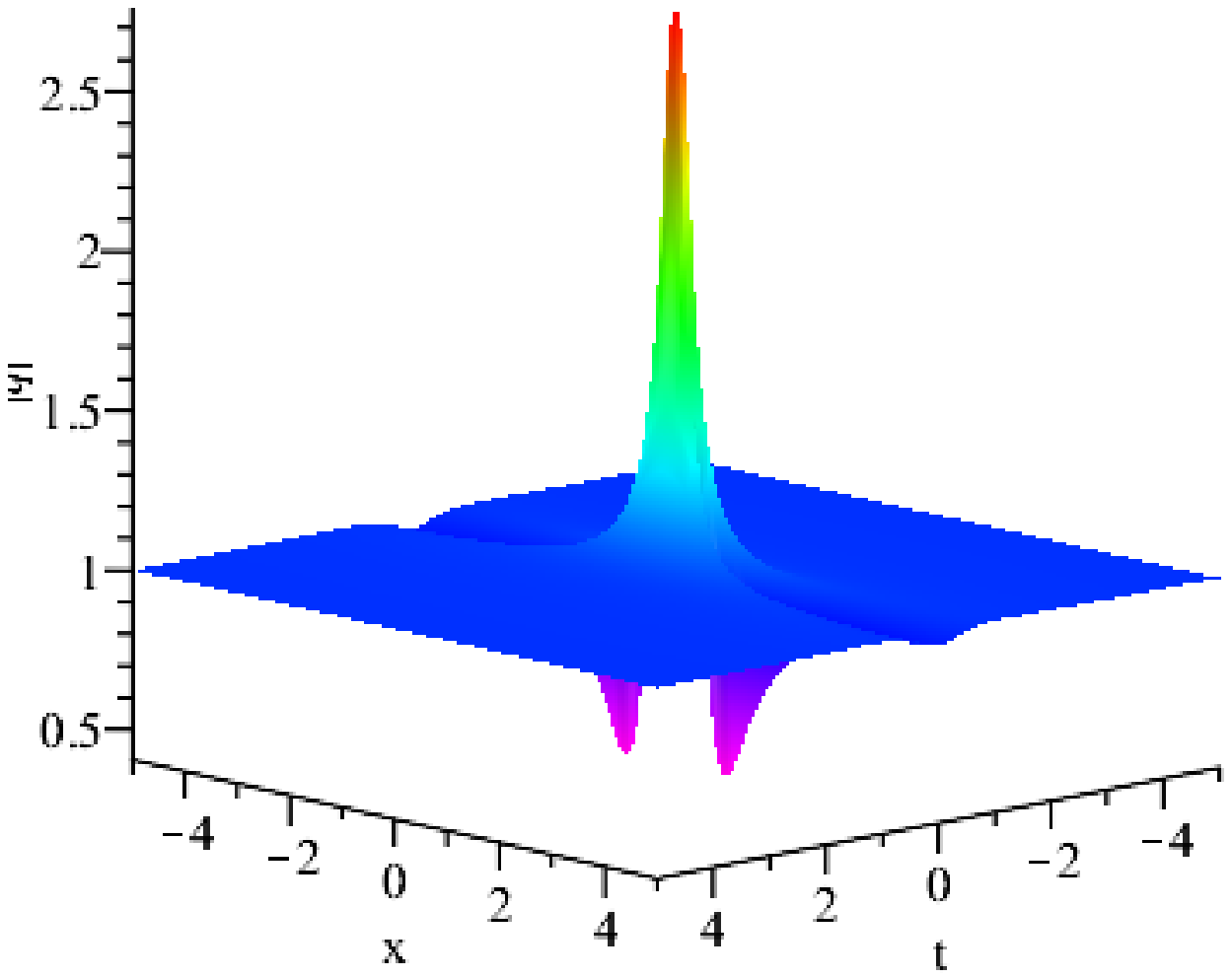}}}
\qquad\qquad\qquad
{\rotatebox{0}{\includegraphics[width=3.6cm,height=3.0cm,angle=0]{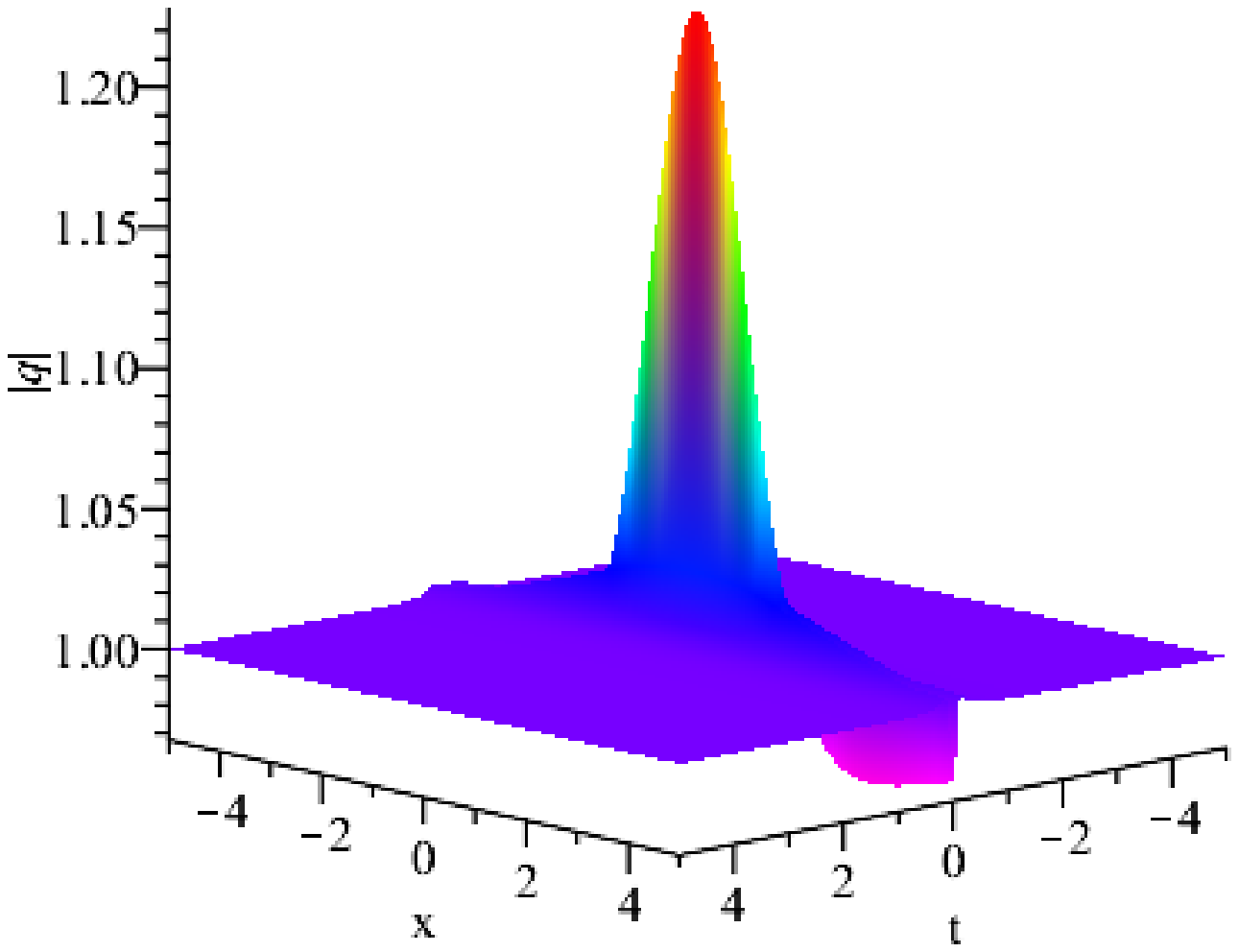}}}\\

\qquad\qquad\qquad$(\textbf{a})\qquad\qquad\qquad\qquad\qquad\qquad\qquad\qquad(\textbf{b})
\qquad\qquad\qquad\qquad\qquad\qquad\qquad\qquad(\textbf{c})$\\

\qquad
{\rotatebox{0}{\includegraphics[width=3.6cm,height=3.0cm,angle=0]{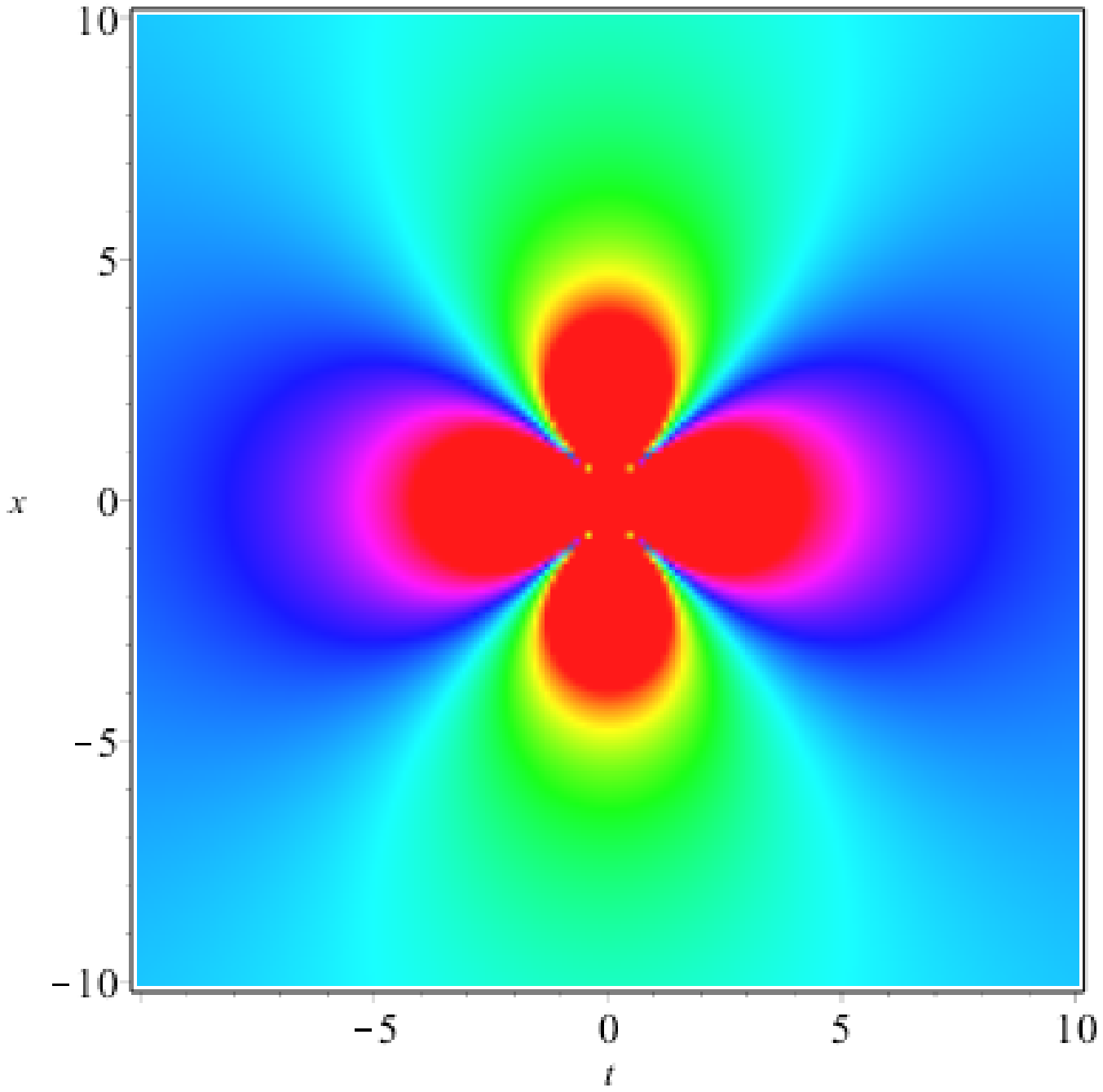}}}
\qquad\qquad\qquad
{\rotatebox{0}{\includegraphics[width=3.6cm,height=3.0cm,angle=0]{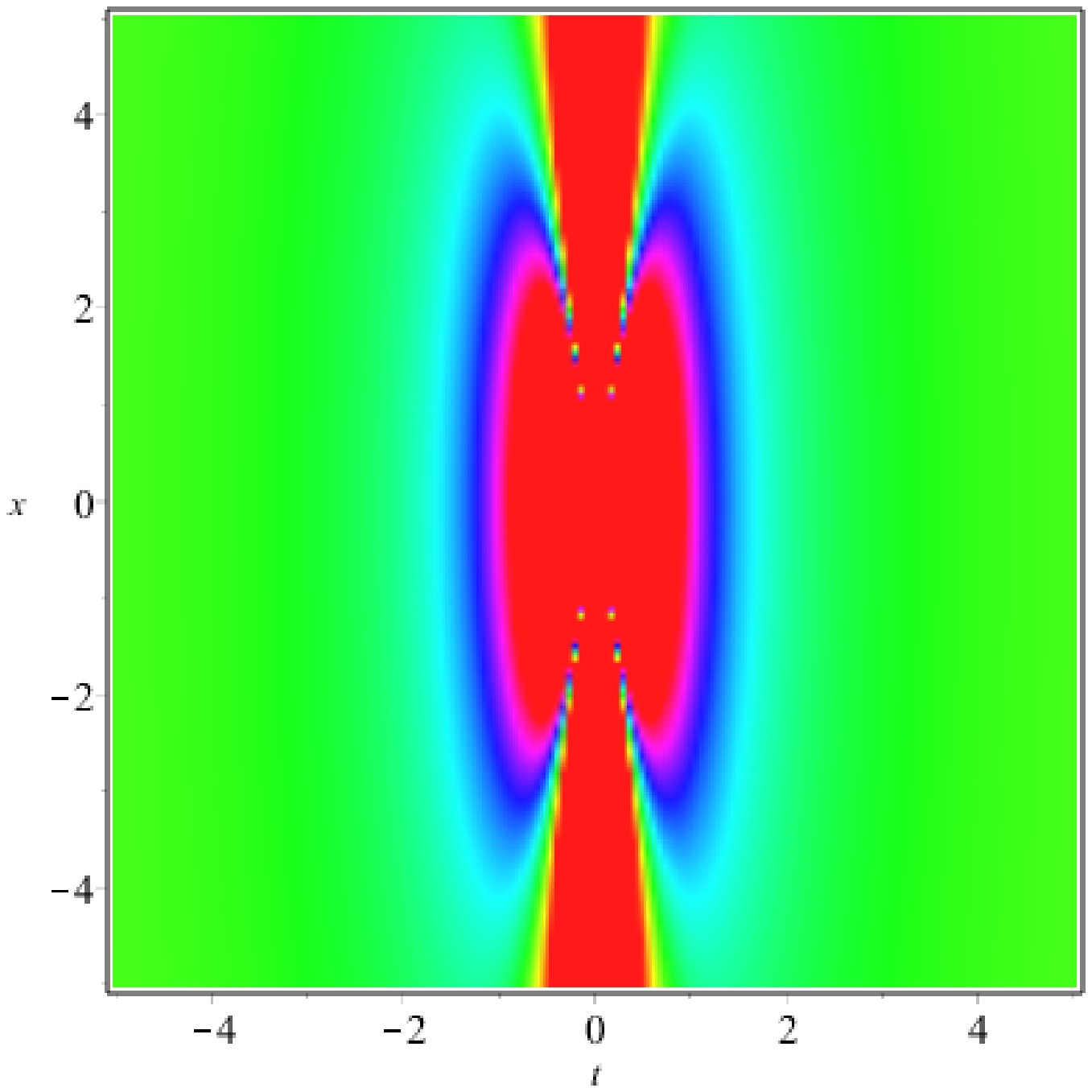}}}
\qquad\qquad\qquad
{\rotatebox{0}{\includegraphics[width=3.6cm,height=3.0cm,angle=0]{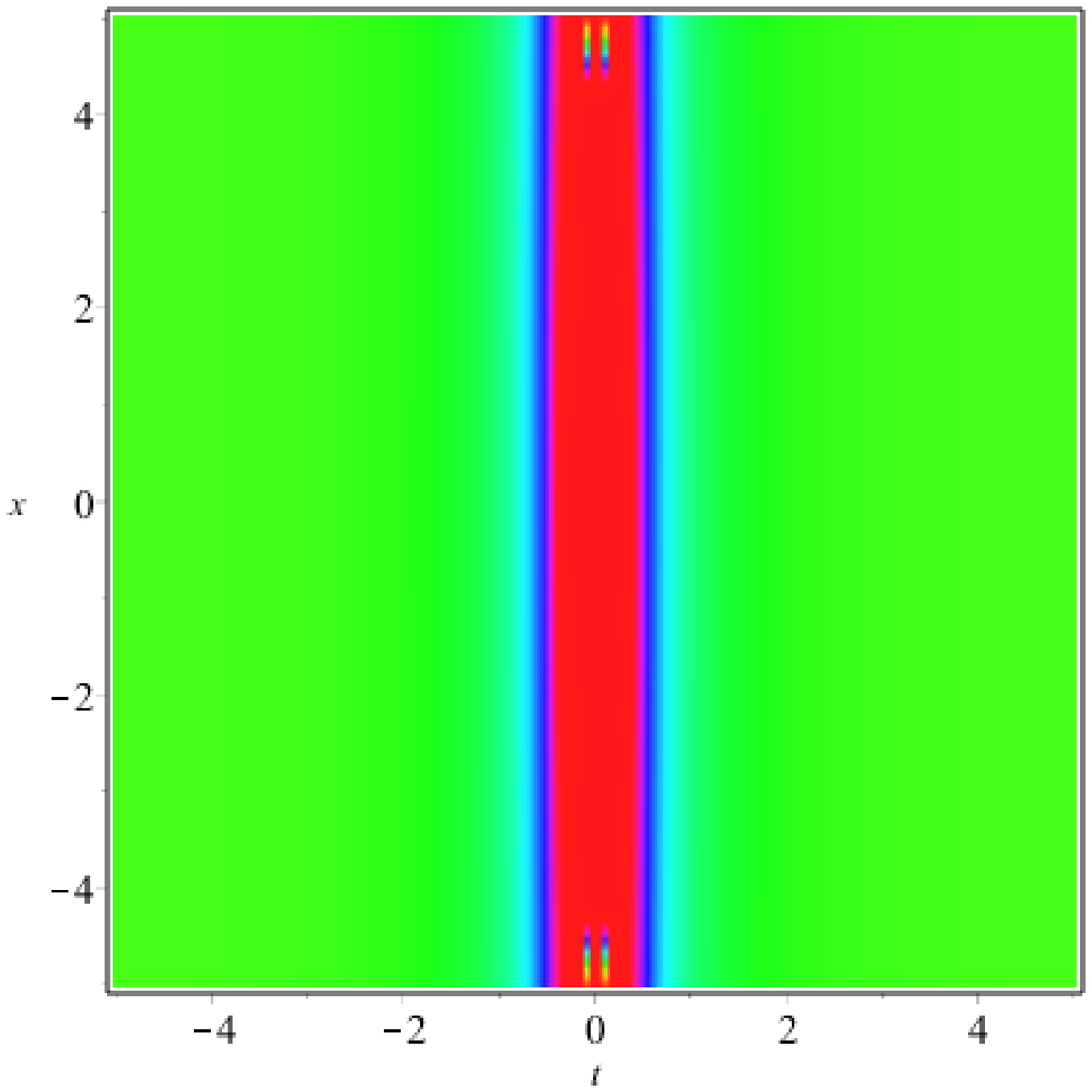}}}\\

\qquad\qquad\qquad$(\textbf{d})\qquad\qquad\qquad\qquad\qquad\qquad\qquad\qquad(\textbf{e})
\qquad\qquad\qquad\qquad\qquad\qquad\qquad\qquad(\textbf{f})$\\
\noindent { \small \textbf{Figure 8.} (Color online) Three dimensional plots and density plots of the first-order rogue wave \eqref{36.2} for Eq.\eqref{1} with the parameters $A=1, \alpha=0$:
$\textbf{(a,d)}$ $\delta=0$ ,
$\textbf{(b,e)}$ $\delta=0.1$,
$\textbf{(c,f)}$ $\delta=1$.}\\

On the other hand, taking $c_{\infty}=(1, i)^{T}$ in \eqref{35}, we will have the second-order rogue wave
\begin{gather}
\widetilde{q}(x, t)=\frac{\Xi_{2}}{\Xi_{1}}e^{i(\alpha x+\beta t)},\label{36.3}
\end{gather}
where
\begin{align}\label{36.4}
&\Xi_{1}=-64A^{6}\theta^{\ast 3}\theta^{3}+96iA^{5}\theta^{\ast 3}\theta'+48A^{4}\theta^{\ast 3}\theta-144A^{4}\theta^{\ast 2}\theta^{2}+48A^{4}\theta^{\ast}\theta^{3}\notag\\
&-72iA^{3}\theta^{\ast}\theta'-108A^{2}\theta^{\ast}\theta-96iA^{5}\theta^{\ast'}\theta^{3}+72iA^{3}\theta^{\ast'}\theta-144A^{4}\theta^{\ast'}\theta'-9,\notag\\
&\Xi_{2}=-64A^{7}\theta^{\ast 3}\theta^{3}-288iA^{5}\theta^{\ast 2}\theta'-192A^{6}\theta^{\ast 3}\theta^{2}-144A^{5}\theta^{\ast 3}\theta-48A^{4}\theta^{\ast 3}\notag\\
&+192A^{6}\theta^{\ast 2}\theta^{3}-216iA^{4}\theta^{\ast '}\theta
+432A^{5}\theta^{\ast 2}\theta^{2}+144A^{4}\theta^{\ast 2}\theta-144A^{5}\theta^{\ast}\theta^{3}\notag\\
&-72iA^{3}\theta'-144A^{4}\theta^{\ast}\theta^{2}+180A^{3}\theta^{\ast}\theta+108A^{2}\theta^{\ast}
-288iA^{5}\theta^{\ast'}\theta^{2}-72iA^{3}\theta^{\ast'}\notag\\
&-96iA^{6}\theta^{\ast'}\theta^{3}-144A^{5}\theta^{\ast'}\theta'+96iA^{6}\theta^{\ast 3}\theta'+48A^{4}\theta^{3}\notag\\
&+216iA^{4}\theta^{\ast}\theta'-108A^{2}\theta-45A,
\end{align}
and the dynamic behavior of the second-order rogue wave is shown in the following pictures.
\\

\qquad
{\rotatebox{0}{\includegraphics[width=3.6cm,height=3.0cm,angle=0]{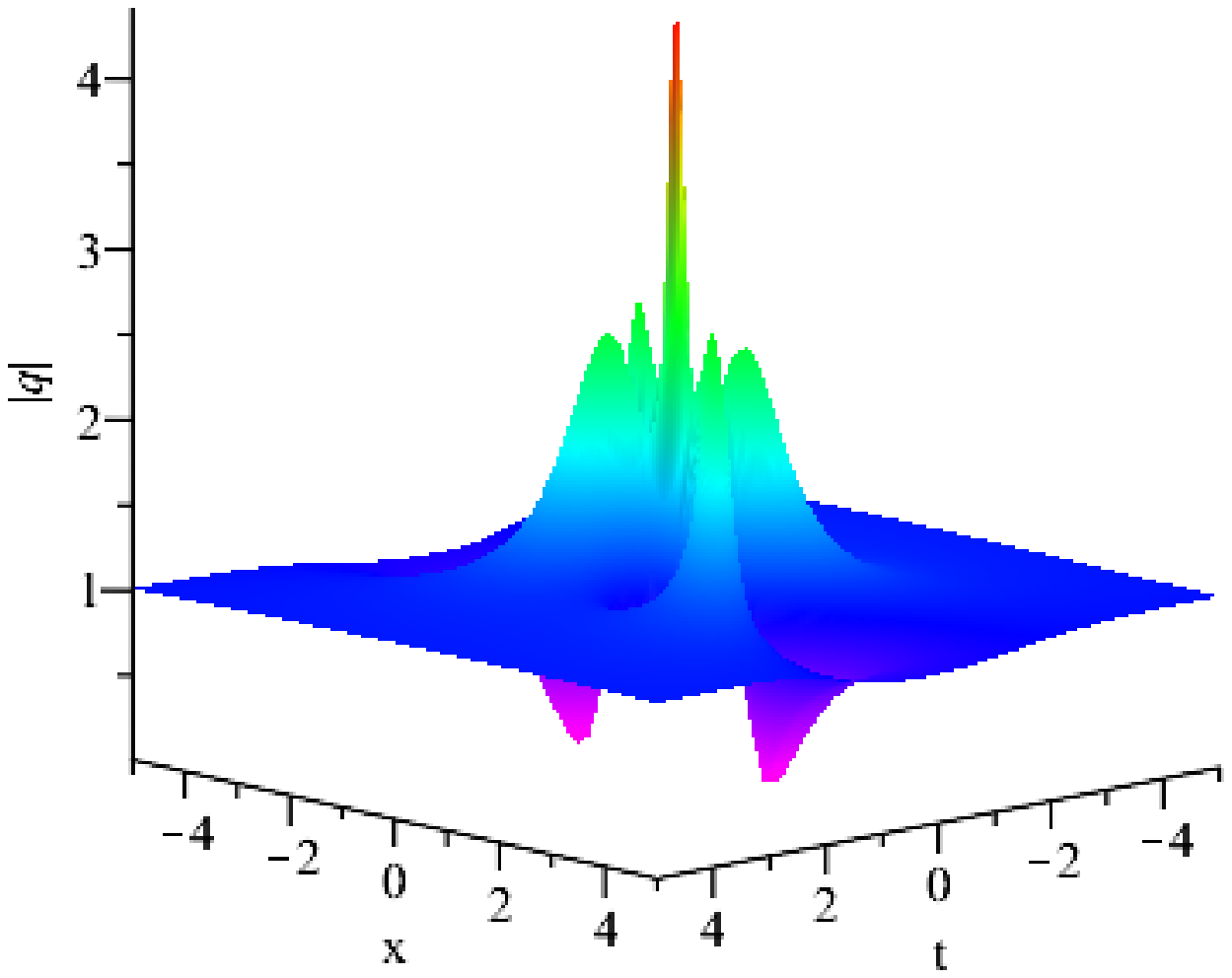}}}
\qquad\qquad\qquad
{\rotatebox{0}{\includegraphics[width=3.6cm,height=3.0cm,angle=0]{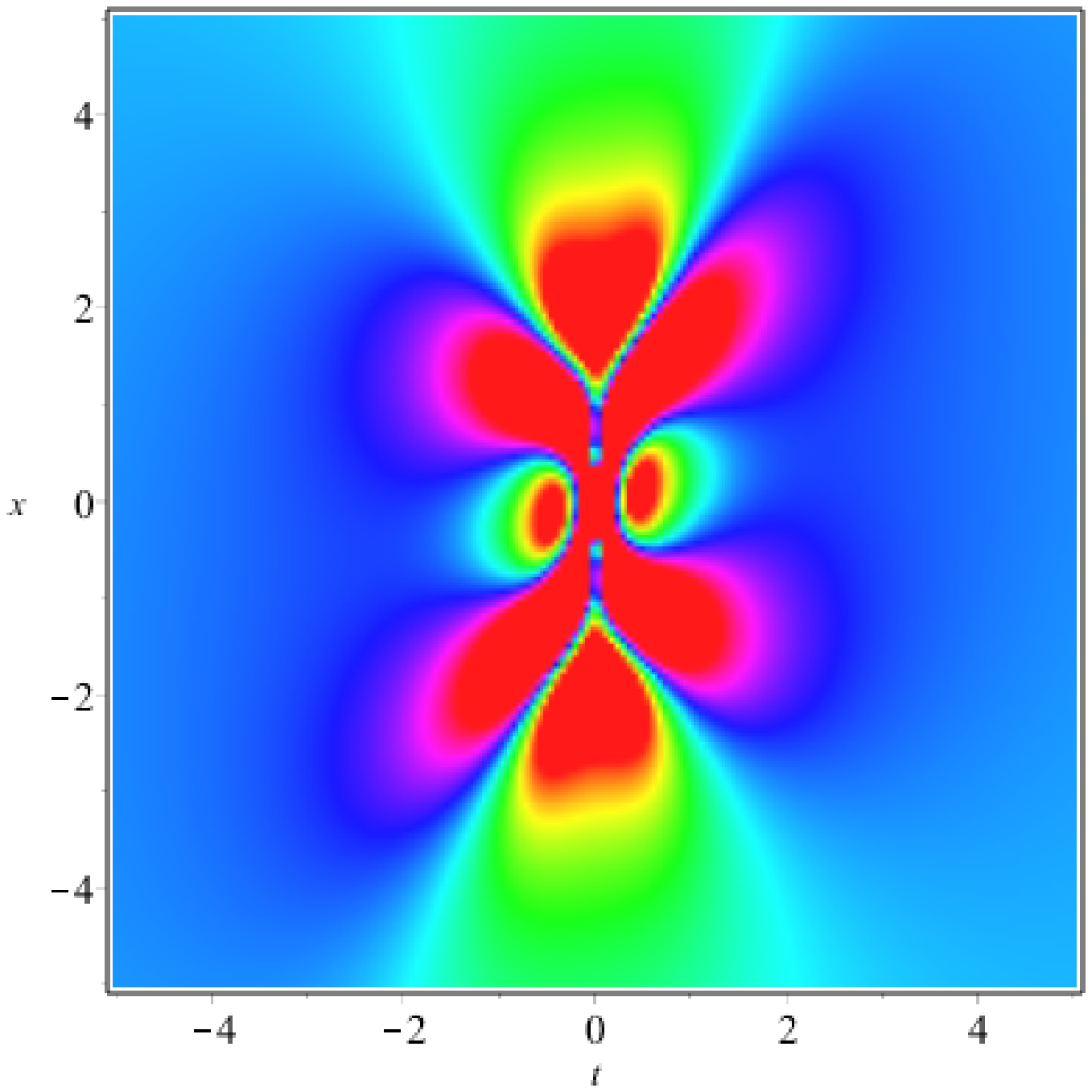}}}
\qquad\qquad\qquad
{\rotatebox{0}{\includegraphics[width=3.6cm,height=3.0cm,angle=0]{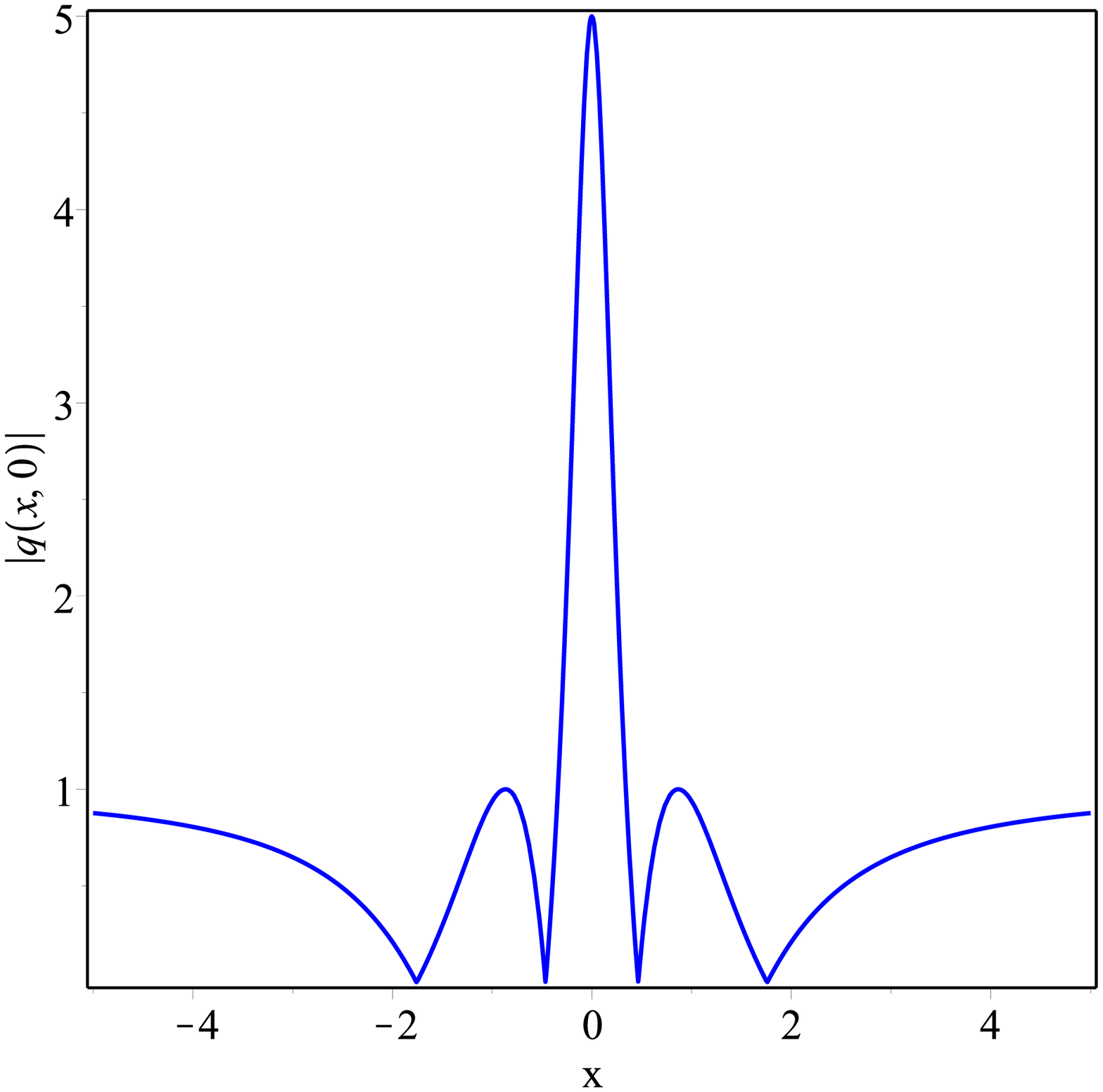}}}\\

\qquad\qquad\qquad$(\textbf{a})\qquad\qquad\qquad\qquad\qquad\qquad\qquad\qquad(\textbf{b})
\qquad\qquad\qquad\qquad\qquad\qquad\qquad\qquad(\textbf{c})$\\
\noindent { \small \textbf{Figure 9.} (Color online) The sceond-order rogue wave \eqref{36.3} for Eq.\eqref{1} with the parameters $A=1, \delta=0.01, \alpha=\frac{1}{10}$. $\textbf{(a)}$ Three dimensional plot;
$\textbf{(b)}$ The density plot;
$\textbf{(c)}$ The wave propagation along the $x$-axis at $t=0$.}\\

\section{Conclusion}
In this paper, we first researched the bound-state soliton  with one higher-order pole for the SONLS equation at ZNCs via using Laurent's series and generalization of the residue theorem. Then, we discussed the corresponding dynamic behavior of the solutions, and indicated that the parameter $\delta$ could decide the waveform and  distance between two waves.  We also  have used the robust inverse scattering transform to derived the breather wave and rogue wave solutions for the SONLS equation with NZBCs. Based on the one-fold DT, we obtained the first order rogue wave solution for $c=(ic, c)^{T}$ and second rogue wave solution for $c_{\infty}=(1, i)^{T}$. Besides, we found that the amplitude of rogue wave only depends on boundary conditions $A$. Furthermore, we also analysed the effect of parameter $\alpha$  and the higher order terms $\delta$  on the wave through choosing different parameter $\alpha$ and $\delta$. The technique shown in this paper can be generalized to some other nonlinear systems and more meaningful phenomena will be presented by certain further research.

\end{document}